\documentclass{aa}  

\usepackage{graphicx}
\usepackage{txfonts}
\RequirePackage{tikz}	
\usetikzlibrary{arrows,backgrounds}
\usetikzlibrary{shapes,arrows}
\usetikzlibrary{positioning}

\usepackage{url}
\usepackage{enumitem}
\usepackage{pifont}

\newcommand\sml{\small}
\usepackage{animate}
\usepackage{xmpmulti}
\newcommand\T{\rule{0pt}{2.6ex}}
\newcommand\B{\rule[-1.2ex]{0pt}{0pt}}

\usepackage{transparent}

\definecolor{Gray}{gray}{0.95}
\usepackage{colortbl}
\definecolor{dca}{rgb}{0.64, 0.0, 0.0}
\definecolor{snow}{rgb}{1.0, 0.98, 0.98}
\usepackage{siunitx}
\usepackage{multirow}
\usepackage{bm}
\usepackage{romannum}

\usepackage{comment} 
\usepackage{soul}

\definecolor{RowColor1}{rgb}{0.8078,0.8627,0.8824}
\definecolor{amber}{rgb}{1.0, 0.75, 0.1}

\definecolor{ceruleanblue}{rgb}{0.16, 0.32, 0.75}
\definecolor{brightcerulean}{rgb}{0.11, 0.67, 0.84}
\usepackage{threeparttable}

\usepackage{xspace}

\newcommand{\lumcgs}{ergs~s$^{-1}$\xspace}

\newcommand\clearrow{\global\let\rowmac\relax}
\clearrow

\begin{document}

   \title{Towards an X-ray inventory of nearby neutron stars}

   \author{A. Vahdat
          \inst{1}\thanks{}
          \and
          B. Posselt\inst{2,3}
          \and 
          A. Santangelo\inst{1}
          \and
          G.G. Pavlov\inst{3}
          }

   \institute{Institut f\"ur Astronomie und Astrophysik, Universit\"at Tübingen, Sand 1, D-72076 T\"ubingen, Germany \\
              \email{arminvahdat@astro.uni-tuebingen.de}
         \and
              Department of Astrophysics, University of Oxford, Denys Wilkinson Building, Keble Road, Oxford OX1 3RH, UK\\
        \and
              Department of Astronomy \& Astrophysics, Pennsylvania State University, 525 Davey Lab, 16802 University Park, PA, USA\\
             }

   \date{Received ----, 2021; accepted ----, 2021}

 \abstract
   {The X-ray emission of neutron stars enables a probe of their temperatures, geometries and magnetospheric properties. The current number of X-ray emitting pulsars is insufficient to rule out
   observational biases that may arise from poorly known distance, age, or location of the neutron stars. One approach to overcome such biases is to create a distance-limited sample with sufficiently deep observations.}
   {With the aim of better sampling of the nearby ($\lesssim$ 2kpc) neutron stars population, we started an \emph{XMM-Newton} survey of pulsars to measure their X-ray fluxes or derive respective constraining upper limits.} 
   {We investigated 14 nearby pulsars for potential X-ray counterparts using different energy bands and detectors. In addition to our new \emph{XMM-Newton} data, we also considered archival data and catalogs.

We discuss source properties and also check for alternative counterparts to the X-ray sources.}
   {In our new \emph{XMM-Newton} data, we found two pulsar counterpart candidates with significance above over $4\sigma$ and one candidate with $3.5\sigma$ by combining EPIC camera detection likelihoods. We also report the detection of potential X-ray counterparts to 8 radio pulsars in the 4XMM-DR10 catalog which have not been reported in the literature.
}
 
   {}


   \maketitle
%

\section{Introduction}
In solitary neutron stars (NS), thermal X-rays are produced by the cooling of NS surfaces and from hot spots/polar caps heated by particles accelerated in the magnetosphere.
On the other hand, the accelerated particles generate non-thermal synchrotron and curvature radiation. Interaction of the pulsar wind with the interstellar medium (ISM) produces shocks that can contribute to the non-thermal X-ray emission. This can contaminate the observed pulsar spectrum considerably if the pulsar and pulsar wind nebula are not spatially resolved.\\

In young, energetic rotation-powered pulsars (RPPs), such as the Crab pulsar, the thermal radiation from the hot neutron star is outshined by non-thermal magnetospheric emission, and the X-ray spectra often have a power-law (PL) shape.
As pulsar ages, thermal  emission becomes more prominent and the spectra can be described by the combination of thermal and non-thermal emission (an example is the Vela pulsar). There are also solitary pulsars that show predominantly thermal X-ray emission, such as Central Compact Objects or the so-called `Magnificent Seven' (e.g., \citealt{Haberl2007Ap, Gotthelf2013}). Lastly, the millisecond pulsars (ms-pulsars) that are assumed to be spun up through accretion from their companion \citep{alpar1982new} usually show both thermal and non-thermal emission components in their X-ray spectra. The small emission radii obtained by spectral fits imply the origin of thermal radiation to be one or two heated polar-caps, whereas the non-thermal emission is presumed to have a magnetospheric origin (e.g., \citealt{1998Zavlin, Forestell2014MNRAS}).\\

X-ray emitting NSs have been identified in surveys such as the ROSAT All Sky Survey (0.1-2.4 keV energy; \citealt{Voges1999A&A}), complemented by X-ray observations with \emph{XMM-Newton} and \emph{Chandra} (e.g., \citealt{becker2009x}).
NSs that were not already known as radio or $\gamma$-ray pulsars were noticed in X-rays either due to signatures of ongoing accretion or simply because they were nearby, young and powerful, or hot. Other NSs were discovered by their variability (outbursts) or activity in other wavelengths, e.g., Soft Gamma-ray Repeaters. 
Many X-ray emitting NSs were found in targeted (deeper) searches based on their radio and $\gamma$-ray properties. 
A few pulsars falling in the field of view (FOV) of pointed observations by X-ray satellites have also been discovered (e.g., \citealt{Pires2009,Prinz2015}). However, many NSs were not covered by sufficiently deep X-ray exposures.
A selection effect cannot be excluded and may lead to a biased view of the known X-ray emitting NS population, which could result in a biased interpretation of their general X-ray properties such as temperatures, spectral indices of non-thermal emission, and X-ray efficiency. Up to now, roughly about 5-7\,\% of the known pulsar population is detected in X-rays.

Pulsars lacking X-ray detection may simply be too faint for the previous surveys, e.g., because they are too far away or too absorbed, or may be indeed ``X-ray quiet''. X-ray monitoring with all-sky surveys such as the on-going eROSITA will provide an important contribution to obtain a flux-limited population of X-ray emitting NSs in the Galaxy. 
A complementary approach that aims for an unbiased coverage of the NS population is to restrict the analyzed sample to a certain distance and increase the number of observed sources in that distance range.
In this study, we will present the first results of an \emph{XMM-Newton} survey aimed at a better sampling of the nearby ($\lesssim 2$\,kpc) NS population.

\begin{table*}[]
\caption{Properties of the ordinary and millisecond pulsars that are investigated with \emph{XMM-Newton} in this study. 
}
\label{tab:source_list}
\setlength{\tabcolsep}{12pt}
	\begin{center}
	\scalebox{0.67}{
	\begin{tabular}{cc ccc ccc ccccc} \hline \T \B
		\multirow{2}{*}{Pulsar}  & \multirow{2}{*}{ObsID} & $t_{\rm PN}$ & $t_{\rm M1}$ & $t_{\rm M2}$ & $b_{\rm PN}$ & $b_{\rm M1}$ & $b_{\rm M2}$ & \it{P} & \it{D}$_1$ & \it{D}$_2$ & log $\dot{E}$ & \multirow{2}{*}{N$_{\rm H,21}$} \T \B \\
	& & (ks) & (ks) & (ks) & (c/s) & (c/s) & (c/s) & (ms) & ({\sml kpc}) & ({\sml kpc}) & ({\sml \it{ergs\,s}$^{-1}$}) &  \\

		\hline  \T \B
		J0711--6830 & 0823030601 & 11.9 (6.1) & 13.8 (9.8) & 13.8 (9.2) & 2.5 (TN) & 0.35 (M) & 0.35 (M)  & 5 &  0.11 & 0.86 & 33.6 & 0.6 \\
		 \T \B
		J0745--5353 & 0823031401 & 18.9 (13.5) & 20.8 (18.4) & 20.8 (16.1) & 0.4 (TN) & 0.2 (TN) & 0.2 (TN) & 215 & 0.57 & 0.25 & 34.0 & 3.8 \\	
		 \T \B
		J0942--5552 & 0823031301 & 18.7 (13.5) & 20.6 (20.2) & 20.6 (20.2) & 0.5 (TN) & 0.35 (TN) & 0.35 (TN) & 664 & 0.30 & 3.8 & 33.5 & 5.6 \\	
		  \T \B
		J0945--4833 & 0823031101 & 17.7 (11.2) & 19.6 (17.5) & 19.6 (17.2) & 0.5 (TN) & 0.2 (TN) & 0.25 (TN) & 331 &  0.35 & 1.51 & 33.7 & 3.0 \\
		 \T \B
		J0954--5430 & 0823030401 & 14.7 (4.7) & 16.6 (6.9) & 16.6 (6.0) & 2.0 (M) & 0.35 (M) & 0.35 (M)  & 473 & 0.43 & 3.96 & 34.2 & 6.2 \\
		  \T \B
		J0957--5432 & 0823031001 & 15.7 (5.6) & 17.6 (17.2) & 17.6 (17.3) & 0.8 (M) & 0.35 (M) & 0.35 (M)  & 203 &  0.45 & 4.33 & 34.0 & 7.0 \\		
		 \T \B
		J1000--5149 & 0823030301 & 7.0 (5.8) & 8.9 (6.2) & 8.9 (8.7) & 0.4 (TN) & 0.35 (M) & 0.35 (M) & 255 &  0.13 & 1.93 & 33.4 & 2.2   \\		

		 \T \B
		J1003--4747 & 0823030201 & 14.7 (9.3) & 16.6 (16.4) & 16.6 (16.3) & 1.0 (TN) & 0.35 (M) & 0.35 (M) & 307 & 0.37 & 2.94 & 34.5 & 3.0  \\		
		  \T \B
		J1017--7156 $^B$ & 0823030701 & 10 (5.3) & 11.9 (10.7) & 11.9 (10.0)  & 0.6 (TN) & 0.2 (M) & 0.25 (M) & 2 & 0.26 & 2.98 & 33.8 & 2.9 \\			
		 \T \B
		J1125--5825 $^B$ & 0823031601 & 18.7 (5.7) & 20.6 (16.9) & 20.6 (16.9) & 0.7 (TN) & 0.35 (M) & 0.35 (M) & 3 & 1.74 & 2.62 & 34.9 & 3.9\\	
		  \T \B
		J1543--5149 $^B$ & 0823030901  & 6.5 (5.3) & 8.4 (8.3) & 8.4 (8.2) &  0.4 (TN) & 0.35 (TN) & 0.35 (TN) & 2 & 1.15 & 2.42 & 34.9 & 1.6 \\
	
		 \T \B
		J1725--0732 & 0823031501 & 15.1 (12.3) & 17.0 (16.7) & 17.0 (16.7) & 0.4 (TN) & 0.35 (M) & 0.35 (M) & 240 &  0.20 & 1.90 & 33.1 & 1.8 \\

		 \T \B
		J1740--3015 & 0823030101 & 4.7 (1.7) & 6.6 (6.0) & 6.6 (5.7) & 1.0 (TN) & 0.35 (M) & 0.35 (M) & 607 & 0.40 & 2.94 &  34.9 & 4.7   \\
		 \T \B
		J1755--0903 & 0823030501 & 4.7 (3.9) & 6.6 (6.5) & 6.6 (6.5) &  0.4 (TN) & 0.35 (M) & 0.35 (M)  & 191 & 0.23 & 1.79 & 33.6 & 2.0 \\
		\hline

	\end{tabular}
	}
	\tablefoot{Pulsar names and their \emph{XMM-Newton} obsid's are given in the first column. The pulsars in binary systems are marked with $^B$. EPIC-pn and EPIC-MOS total and GTI-filtered exposure times (given in parentheses) are displayed in the second column. 
The GTI filter for the background light curve and the corresponding filters are listed in third column for each EPIC detector. ``TN'' is thin filter and ``M'' is medium filter. Source properties are noted in the last column. P is the spin period in seconds, D$_1$ is the best estimate of the pulsar distance in kpc according to ATNF pulsar catalog which uses the \emph{YMW+17} DM-based distance as default, {\it D}$_2$ is the distance in kpc based on the NE2001 model, log $\dot{E}$ is the common logarithm of the spin-down energy loss rate in ergs\,s$^{-1}$ and $N_{\rm H}$ is the hydrogen column density in unit of $ 10^{21}$ cm$^{-2}$ estimated from the DM as outlined in \S\ref{ss:sa}.}
	\end{center}
\end{table*}

Although thermal and non-thermal emission can co-exist during the lifespan of a NS, the dominant type of emission depends on its physical properties such as age and overall energy budget.
One key question at the origin of X-ray emission from pulsars is to figure out how much of the available energy budget is converted into X-rays, and how this changes over the NS lifespan. 

There is a general correlation between the X-ray luminosity ($L_{\text X}$) from non-thermal and thermal emission (from polar caps) and spin-down luminosity ($\dot{E}$) of NSs, but the X-ray efficiency ($\eta_{\rm X}$ = $L_{\text X}/\dot{E}$) shows a large variation ($\eta_{\rm X} \sim 10^{-6} - 10^{-2}$). The reported spread of $\eta_{\rm X}$ can be related to the choice of energy range, spectral model, and consideration of extended emission \citep{becker1997x, possenti2002re, li2008nonthermal, kargaltsev2012chandra}.  For example, the fluxes in the soft X-ray range are more sensitive to the uncertainties due to the interstellar absorption which may cause a biased view towards the population of thermally dominated NSs.

Furthermore, it was pointed out by several authors that the X-ray emission becomes more efficient at $\dot{E} \lesssim 10^{34}-10^{35}$ erg\,s$^{-1}$. This could support polar cap and pair cascade model predictions indicating an increase in electron-positron pair heating luminosity \citep{harding2011pulsar, Harding2001ApJ} or it could reflect the presence of the dominant non-dipole magnetic field \citep{kisaka2017efficiency}. However, there are too few pulsars with $\dot{E} \lesssim 10^{34}$ ergs\,s$^{-1}$ to constrain theoretical models. Different sample selections may lead to different X-ray efficiencies. For instance, in middle-aged pulsars the X-ray efficiencies, $\eta_{\rm X} \sim 10^{-3} - 10^{-2}$, seem to be higher than in younger ones, $\eta_{\rm X} \sim 10^{-5} - 10^{-3}$ \citep{kargaltsev2012chandra, posselt2012chandra}.\\
The number of X-ray detected nearby pulsars is currently insufficient to exclude selection biases towards their observational properties. We approach this problem by conducting a survey of the nearby ($\lesssim$ 2 kpc) NS population. In \S\ref{oaa}, we describe our source selection strategy and data analysis steps for our survey, including the consideration of nearby multiwavelength sources as alternative X-ray counterparts. In \S\ref{sec:results}, we present initial results of the survey where we provide significance and upper limits for the observed sources.
\S\ref{sec:hiddenp} covers the details of our additional serendipitous archival candidates study. Finally, \S\ref{sec:discussion} discusses the X-ray efficiency of sources, the effect of different distance estimates, and 
implications. 

\section{Observations and analyses}\label{oaa}
\subsection{Source selection and observations}

We used the ATNF catalog (Manchester et al. \citeyear{Manchester2005AJ}, V1.57) to select all known nearby ($\lesssim$ 2 kpc) NSs that had no X-ray coverage in the field of view of archived \emph{XMM-Newton} pointing (search radius $13\arcmin$ around pulsar position) and \emph{Chandra} pointing (search radius $10\arcmin$) observations.
We choose 2 kpc as a cut-off threshold for the distance of the pulsars. Above this value, the absorption effects become much more problematic for thermal emission, making it harder to detect NS where this type of emission dominates \citep[e.g., the Magnificent Seven,][]{Haberl2004}.\\

We used distance estimates from the ATNF which are typically based on the dispersion measure (DM) from radio timing observations, and the electron density model by \cite{Yao2017ApJ} (\emph{YMW+17}).
Some of these estimates may differ up by to an order of magnitude from the results based on the electron density model by \citet{Cordes2002astro} (\emph{NE2001}). We also list these distances for completeness in Table \ref{tab:source_list} and discuss distance differences of these estimates in \S\ref{ssec:distance}. Using the pulsar's $\dot{E}$, distance $D$, and average expected hydrogen column density (${N}_{\rm H}$) calculated from the DM, we assumed a preliminary X-ray efficiency of $10^{-4}$  to estimate expected absorbed X-ray fluxes. We started our survey with those pulsars for which a detection seemed most likely in a short exposure time. 14 pulsars were observed with the \emph{XMM-Newton} EPIC instrument in 2018, with 5-20 ks exposures (\emph{Program ID: 082303}) with \emph{Thin} and \emph{Medium} filters. Here we present the results of the first year of the survey. The list of the so far observed survey sources, their exposure times, and selected observational properties are presented in Table \ref{tab:source_list}.

\subsection{Source detection optimization}\label{ss:so}
To optimize the choice of our energy bands for source detection, we analyzed the detection likelihoods of known X-ray-detected pulsars in 4XMM-DR10. We found that the energy bands 0.3-2.0 keV and 1.0-4.5 keV have the highest number of detections based on their maximum likelihood (ML) distributions compared to other bands in soft and mid-X-ray energies. We then analyzed a few pulsars with already known X-ray detections using the archival data to test these values and also checking for other broader or narrower bands. We decided to add the following energy ranges for our analyses: 0.3-1.2 keV, 0.3-1.5 keV, and 0.3-4.5 keV which provided the highest detection significance defined as:

\begin{equation}
\label{eq1}
S = \frac{N_{\rm s}}{\hat{\sigma}(N_{\rm s})} = \frac{N_{\rm T} - \alpha N_{\rm bgr}}{\sqrt{N_{\rm T} + \alpha^2 N_{\rm bgr}}} 
\end{equation}

\noindent where $N_{\rm s}$ is the number of background subtracted source counts, $N_{\rm T}$ and $N_{\rm bgr}$ are total counts within the source and background regions, and $\alpha$ is ratio of source to background region.\footnote{Although we list the conservative S (formula~\ref{eq1}), we note that an alternate significance definition by \citet{LiMa1983}, \protect\\ $S = \sqrt{2} \left\{ N_{\rm T}\ln{\left[ \frac{\alpha + 1}{\alpha} 
\left(\frac{N_{\rm T}}{N_{\rm T}+N_{\rm bgr}}\right)\right]} \right.  \left. + N_{\rm bgr} \ln{ \left[ (\alpha +1) \frac{N_{\rm bgr}}{N_{\rm T}+N_{\rm bgr}} \right] } \right\}^{1/2}$ would result in a factor of $\sim$ 1.2-1.4 higher significance.}

\subsection{Data reduction and analyses}\label{ss:dr}
All data have been reduced with \emph{XMM-Newton} Science Analysis Software (SAS, V18.0.0.) applying standard tasks. We filtered with {\ttfamily FLAG==0} to avoid CCD gaps and bad pixels. We have performed a good time interval (GTI) screening to identify and remove flares caused by energetic protons. Since our short observations were impacted by many background flares, standard background flare screening would have removed often more than half of the exposure time. Therefore, we allowed a stronger background contribution in some observations to maximize potential photon numbers from our target. The full list of GTI rates used for observations are presented in Table\,\ref{tab:source_list}.  
We tested several GTI filters derived from 100 s binned light-curves of events with energies above 10 keV. For each GTI-filtered event file, we generated images and exposure maps in 5 energy bands as specified above. We then employed the SAS-tool {\ttfamily eregionanalyse}  to optimize the aperture that maximizes the source to background count ratio. The range of the optimal extraction radii provided by the routine varied between 10$\arcsec$--15$\arcsec$. We allowed {\ttfamily eregionanalyse} to center on the position that is up to $3\sigma$ from the radio pulsar position\footnote{For positions, proper motions, and parallaxes we used the values listed in the ATNF catalog V1.64}. We also applied a proper motion correction in our $3\sigma$  position uncertainty circle (if available). Only one source (PSR J0745$-$5353) in our survey has a  noticeable proper motion ($\mu_{\alpha}$ =-60$\pm$10 mas/year, $\mu_{\delta}$ =50$\pm$10 mas/year, Parthasarathy et al. \citeyear{parthasarathy2019timing}). 
We obtained the background-subtracted source count rates as well as the $3\sigma$ statistical upper limits for each X-ray NS candidate covered by the individual EPIC instruments.\\ 

In order to obtain combined EPIC count rates and significance, we used the {\ttfamily emldetect} task, and performed simultaneous ML point spread function (PSF) fits to the source count distributions of the EPIC-pn and EPIC-MOS events. The respective values and upper limit values are given in Table \ref{tab:source_list2}. {\ttfamily emldetect} also provided the statistical positional error for X-ray sources in our survey and by combining it with the systematic error of $\ang{;;1.5}$ \citep{Webb2020}, we obtained the total positional uncertainties to search for multiwavelength sources in the proximity of X-ray NS candidates.\\ 

For sources with a combined net count number above 100 in the EPIC instrument, we also performed a spectral fit. 
PSR J1831--0952 was the only source in our list that satisfied this criterion. For this source we generated source and background spectrum from the extraction region provided by {\ttfamily eregionanalyse}. We produced redistribution matrices and effective area files using the usual SAS tasks {\ttfamily rmfgen} and {\ttfamily arfgen}. We then used the SAS task {\ttfamily specgroup} to group the source counts of each spectrum with $\geqslant$ 15 per bin. We repeated the procedure with 5 cts binning and \textit{cstat} test statistic and obtained similar results, therefore we only report the former. We provide the spectral fit results in \S\ref{ssec:j1831}.

\begin{table*}
\caption[]{Properties of the selected ordinary and millisecond pulsars that are investigated in this study.
\label{tab:source_list2}}
\setlength{\tabcolsep}{12pt}
	\begin{center}
	\scalebox{0.65}{
	\begin{tabular}{c c c ccc ccc ccc ccc} \hline \T \B
	\multirow{3}{*}{\large{Pulsar}}  & \multirow{3}{*}{\large{ObsID}} & 
	\multirow{2}{*}{\large{\textbf{$\Delta_{\rm X-R}$}}} &
	\multicolumn{3}{c}{PN$_{\rm \,0.3-2\,keV}$} &
	\multicolumn{3}{c}{MOS1$_{\rm \,0.3-2\,keV}$} &
	\multicolumn{3}{c}{MOS2$_{\rm \,0.3-2\,keV}$} & \multicolumn{3}{c}{Combined \hspace{0.2cm} EPIC} \T \B  \\
	 & & & $N_s$ & $S$ & UL$^1$ & $N_s$ & $S$ & UL$^1$ & $N_s$ & $S$ & UL$^1$ & $N_s$ & $S$ & ML \T \B \\
	 & & (arcsec) & (counts) & ($\sigma$) & (c/s) & (counts) & ($\sigma$) & (c/s) & (counts) & ($\sigma$) & (c/s) & (counts & ($\sigma$) & \T \B \\

		\hline \T \B
		J0711--6830 & 0823030601 & 1.7 & 
        23.5$\pm$13.0 & 1.8 & 0.0080 & 
        $--$ & $--$ & 0.0005 &
        11.3$\pm$4.7 & 2.4 & 0.0018 & 54.8$\pm$15.7 & 3.5$^1$ & 14.9\\

		 \T \B
		J0745--5353 & 0823031401 & 1.5 & 
        27.2$\pm$7.6 & 3.6 & $--$ &
        6.2$\pm$4.0 & 1.6 & 0.0006 &
        8.1$\pm$4.3 & 1.9 & 0.0007 & 39.6$\pm$9.2 & 4.3$^2$ & 16.2\\
		 \T \B
		J0942--5552 & 0823031301 & 9.7 & 
		31.7$\pm$10.3 & 3.1 & $--$ &
		$--$ & $--$ & 0.0004 & 
		$--$ & $--$ & 0.0006 & $--$ & $--$ & $--$\\

		 \T \B
		J0945--4833 & 0823031101 & 2.3 & 
        14.8$\pm$7.2 & 2.2 & 0.0021 &
        3.3$\pm$5.1 & $--$ & 0.0007 &
        $--$ & $--$ & 0.0004 & $--$ & $--$ & $--$\\

		 \T \B
		J0954--5430 & 0823030401 & 1.2 & 
        4.4$\pm$4.2 & 1.1 & 0.0021 & 
        1.7$\pm$3.2 & $--$ & 0.0009 & 
        2.5$\pm$2.9 & $--$ & 0.0012 & $--$ & $--$ & $--$\\

		 \T \B
		J0957--5432 & 0823031001 & 8.5  &
		9.0$\pm$5.8 & 1.6 & 0.0024 & 8.3$\pm$5.6 & 1.5 & 0.0008 & 
		$--$ & $--$ & 0.0003 & $--$ & $--$ & $--$\\

		 \T \B
		J1000--5149 & 0823030301 & 2.7 & 
		$--$ & $--$ & 0.0007 &   
		$--$ & $--$ & 0.0005  &   
		$--$ & $--$ & 0.0010 & $--$ & $--$ & $--$ \\

		  \T \B
		J1003--4747 & 0823030201 & 2.1 & 
        10.5$\pm$8.4 & 1.3 & 0.0019 &  
        3.4$\pm$3.8 & $--$ & 0.0005 &
        $--$ & $--$ & 0.0002 & $--$ & $--$ & $--$\\
        
		 \T \B
		J1017--7156 & 0823030701 & 5.3 & 
        1.6$\pm$3.9 & $--$ & 0.0015 &
        $--$ & $--$ & 0.0002 &
        $--$ & $--$ & 0.0003 & $--$  & $--$\\        
		 \T \B
		J1125--5825 & 0823031601 & 0.6 & 
		18.8$\pm$6.7 & 2.8 & 0.0038 & 17.4$\pm$6.3 & 2.8 & 0.0012 &
		10.3$\pm$5.2 & 2.0 & 0.0009 & 47.4$\pm$10.2 & 4.7$^2$ & 20.9\\        
        
		 \T \B
		J1543--5149 & 0823030901  & 1.2 & 
		4.0$\pm$4.0 & 1.0 & 0.0018 & 3.8$\pm$3.6 & 1.2 & 0.0011 & 
		$--$ & $--$ & 0.0007 & $--$ & $--$ & $--$\\        
        
		 \T \B
		J1725--0732 & 0823031501 & 4.4 & 
		3.3$\pm$3.8 & $--$ & 0.0006 & 1.6$\pm$2.7 & $--$ & 0.0004 &
		$--$ & $--$ & 0.0003 & $--$ & $--$ & $--$\\

		 \T \B
		J1740--3015 & 0823030101 &  7.8 & 
        6.4$\pm$3.5 & 1.8 & 0.0064 &
        $--$ & $--$ & 0.0008 &
        $--$ & $--$ & 0.0008 & $--$ & $--$ & $--$\\

		 \T \B
		J1755--0903 & 0823030501 & 0.3 & 
        6.2$\pm$4.1 & 1.5 & 0.0031 & 
        $--$ & $--$ & 0.0006 &
        $--$ & $--$ & 0.0009 & $--$ & $--$ & $--$ \\ \hline

	\end{tabular}
	}
	\end{center}
		\tablefoot{The $\Delta_{\rm X-R}$ column displays angular separation in arcseconds between the pulsar's radio timing position and the centroid of the nearest X-ray source. Background-subtracted counts $N_s$ at the optimized X-ray source position and their corresponding significance $S$ according to Equation (1) are listed for each detector, as well as the ML for the detections considering all EPIC instruments together. The $3\sigma$ upper limits (UL) for individual instruments are Bayesian upper limits according to \cite{Nousek1991ApJ}. The ML values are given in 0.3-2.0 keV for PSRs J0745--5353 and J1125--5825 and in 0.3-4.5 keV for PSR J0711--6830.}
\end{table*}

   \begin{figure*}
   \centering
   \includegraphics[width=0.326\textwidth]{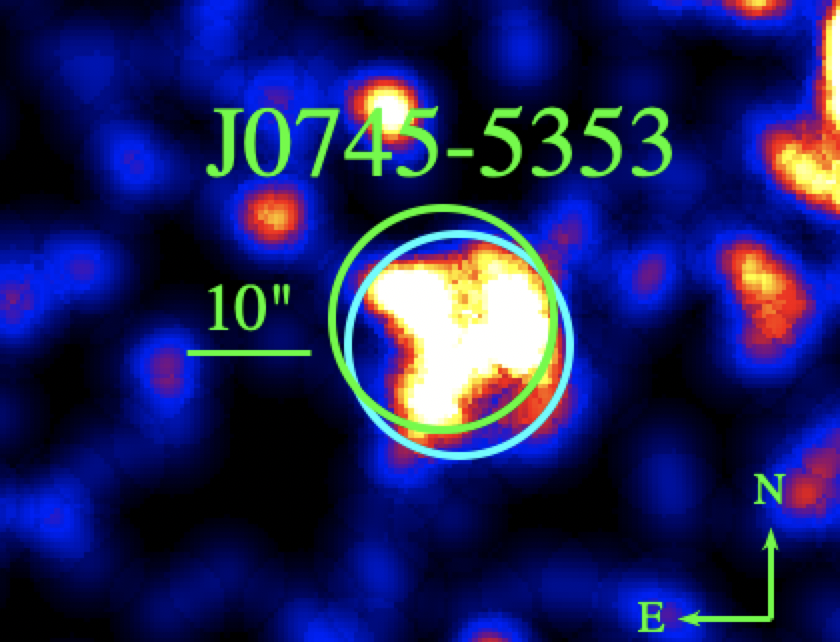}
   \includegraphics[width=0.326\textwidth]{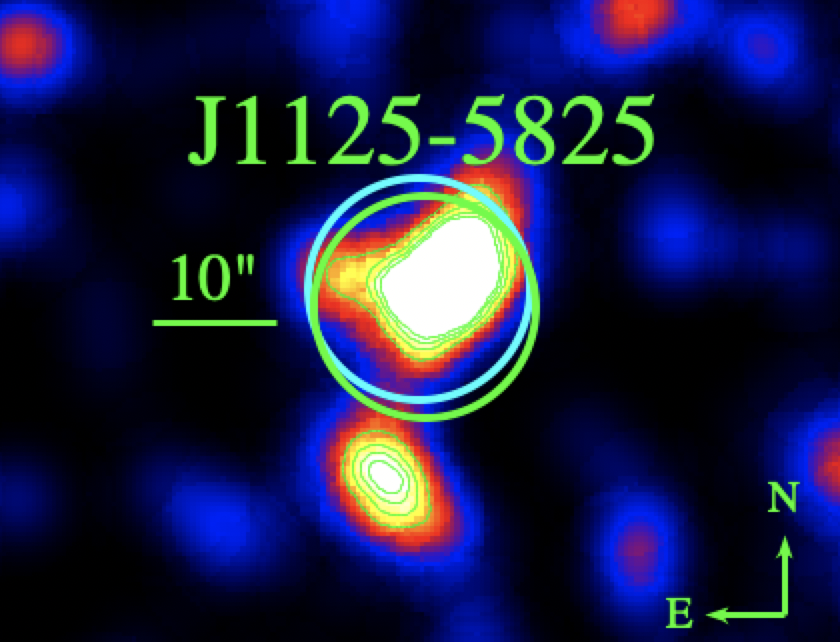}
   \includegraphics[width=0.33\textwidth]{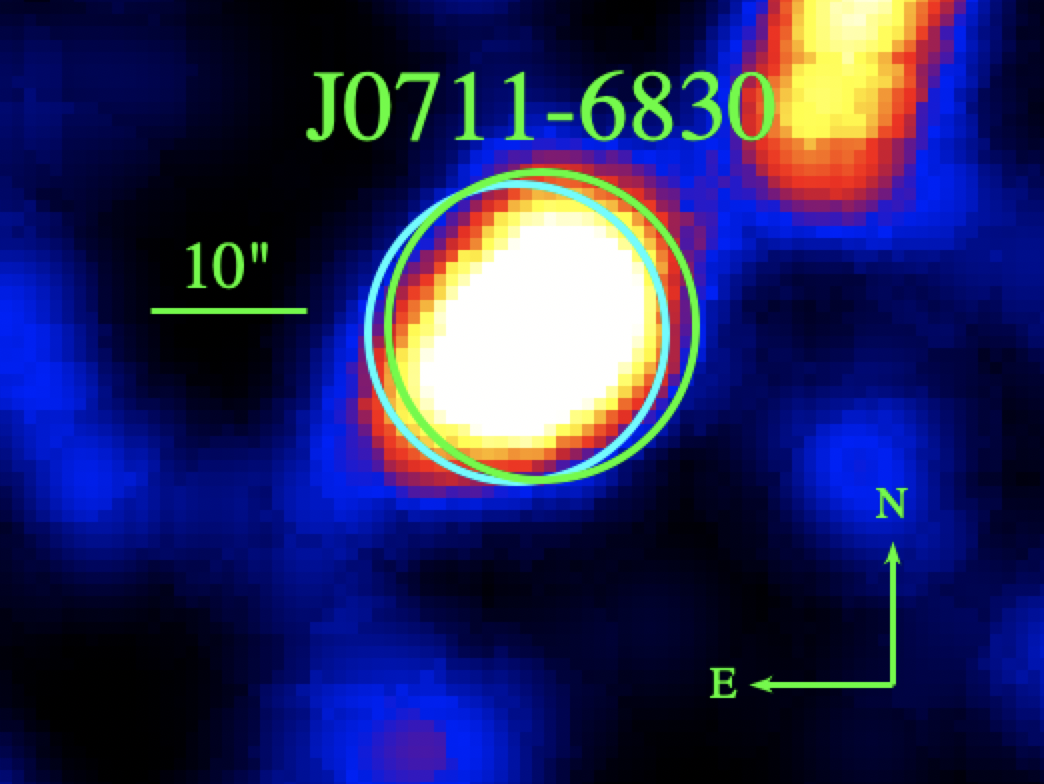}
   \caption{X-ray counterpart candidates in 0.3-2.0 keV as observed with the EPIC-pn and EPIC-MOS2 (PSR J0711-6830).
            The green circles are centered at the radio timing position of the pulsars, the cyan circles are centered at the position where the X-ray source has the highest significance. The images
have pixel scales of $0.4\arcsec$\,, and are smoothed with Gaussian length scales $\sigma$ =3.4$\arcsec$ (for PSR\,J1125--5825), and $\sigma$ =2.4$\arcsec$ (for PSRs\,J0745--5353 and J0711--6830). The angular separations between X-ray and radio positions are given in Table \ref{tab:source_list2}. 
            }
              \label{fig:survey}%
    \end{figure*}

\subsection{Flux estimates}\label{ss:sa}

Since there are typically not enough counts to produce meaningful spectral fits for our sources, we had to assume a spectrum to convert the photon flux to energy flux. Ordinary and ms-pulsars can have both thermal and non-thermal emission components. Aiming to represent two extremes, we decided to convert count rates to fluxes assuming two different spectral models, namely a power-law (PL) with a photon index of 1.7 (e.g., \citealt{becker2009x}) and a blackbody (BB) with {\it kT} = 300 eV (e.g., \citealt{Forestell2014MNRAS}). We used {\ttfamily  PIMMS v4.11} for this conversion. To account for absorption, we estimated the  ${N}_{\rm H}$ values using the formula  $N_\mathrm{H}\rm\;(10^{20}\,cm^{-2})=
0.30^{+0.13}_{-0.09}\;DM$ from \citet{He2013ApJ}.

For both the BB and PL models, we chose the input energy ranges of 0.3-2.0 keV and 1.0-4.5 keV respectively to convert EPIC-pn count rates to absorbed and unabsorbed fluxes. For the PL model, we chose an output energy range of 1-10 keV for easier comparison with the literature. Since no good parallax or alternative distance estimates were available for our pulsars, we used the DM-based distances \citep{Yao2017ApJ} to convert unabsorbed fluxes and upper limits to isotropic X-ray luminosities $L_{\rm 1-10\,keV} = 4 \pi D^2 F^{\text{unabs}}$.

\subsection{Are there alternatives to explain the X-ray sources?}\label{ssec:mw1}

We used the $3\sigma$ position uncertainty radius to check for nearby optical or infrared sources that could be the origin of, or contribute to the observed X-ray emission. We used the \emph{Gaia}-$\rm{EDR3}$ catalog  \citep{2020arXivRiello}, $\rm{2MASS}$ \citep{2MASS06}, $\rm{WISE}$ \citep{WISE10}, $\rm{NOMAD}$ \citep{Zacharias2004AAS} and $\rm{TESS}$ Input Catalog \citep{Stassun2018AJ}.
To test the hypothesis of a different counterpart,
we assumed that all the X-ray flux could come from an optical/NIR source, and investigated the resulting X-ray/optical/NIR flux ratios, color-color, and color-magnitude diagrams, as well as proper motions and distances if known. We used this information to classify each optical/NIR source based on the studies by: 
\cite{Maccacaro1988ApJ}, \cite{Covey2007AJ} and \cite{Jones2019}.

We obtained visual magnitudes ($m_V$) by converting the \emph{Gaia}-$\rm{EDR3}$ passbands and colors using Landolt standard stars observed with \emph{Gaia} \citep{2020arXivRiello}.  
We interpret the obtained X-ray to optical flux ratios by using the classification scheme by \citet{Maccacaro1988ApJ} to determine whether the optical source has a high likelihood of being e.g., an AGN or a main sequence star. Objects such as galaxies, and AGNs might not follow the \emph{Gaia} photometry conversion relations \citep{2020arXivRiello}. However, we also used astrometric properties such as distances and proper motions from \emph{Gaia}-$\rm{EDR3}$ to verify a stellar origin. 
In the absence of $\rm{2MASS}$ counterparts, we converted \emph{Gaia} colors to $\rm{SDSS}$ colors \citep{2020arXivRiello} for classification shown in Table 3 of  \cite{Covey2007AJ}.  
\cite{Jones2019} did an empirical classification of \emph{Gaia}-$\rm{DR2}$ objects (with G $\geqslant$ 14.5 magnitudes) using only \emph{Gaia}-$\rm{DR2}$ data. They used a probabilistic classifier to catalogue sources into star, quasar or galaxy based on their \emph{Gaia} colors, magnitudes, and astrometric features such as parallax and proper motion. We 
took their results into account to classify the potential multiwavelength counterparts in our list.

Finally, we considered the classification  into point sources (stars) and extended sources (e.g., galaxies) provided by the $\rm{TESS}$ input catalog. We found 10 potential optical/NIR counterparts within the  $3\sigma$ position uncertainty regions of 6 of our X-ray candidates (including 5 archival X-ray sources, see section \S\ref{sec:hiddenp}). In order to determine whether these optical/NIR sources can be excluded  as counterparts of the X-ray emission, we evaluated the respective results of the mentioned studies  (Table \ref{gaia}).

\section{Results}\label{sec:results}
Among the five energy bands mentioned in \S\ref{ss:so}, 0.3$-$2.0 keV provided the highest overall detection significance in the EPIC instruments for our survey sources. We, therefore, present the results for this band. We note that the sensitivity loss due to background flaring events (up to 65\% per cent exposure time loss) was larger than anticipated for our sources.

In the framework of our program, we detected three X-ray point sources with at least $3.5\sigma$ sufficiently close to the radio positions of pulsars J0711--6830, J0745--5353, and J1125--5825 to investigate a possible association. The exposure times, used distances, and source properties are listed in Table \ref{tab:source_list}. The angular separation between radio and X-ray position of these pulsars and their EPIC-pn and EPIC-MOS significance are given in Table \ref{tab:source_list2}. 
The corresponding EPIC images of the X-ray counterpart candidates to the three pulsars are displayed in Figure \ref{fig:survey}. Their estimated flux and luminosity values are given in Table \ref{tab:source_list3}.

\subsection*{PSR J0711--6830}

The X-ray counterpart candidate is detected $\ang{;;1.7}$ off from the radio position of the pulsar. The source may be an example of strong non-thermal emission as the inclusion of higher energies resulted in a detection in contrast to the narrower soft energy band that had not enough counts. However, detailed inspection of the data revealed an irregular shape of the count distribution in EPIC-MOS2, a noise-like appearance in EPIC-pn, and the absence of the source in EPIC-MOS1, raising doubts about the reality of the detection of this source.

\subsection*{PSR J0942--5552}

For this X-ray source, the location with the highest significance according to {\ttfamily eregionanalyse}, is $9.7\arcsec$ away from the radio pulsar position. However, the lack of detection in EPIC-MOS1/MOS2 and the high angular separation between X-ray and radio position prevents us from claiming a match. We also noticed a nearby (unknown) extended X-ray source, but the angular separation of $\approx 25\arcsec$ between its emission peak and the radio pulsar position indicates that an association with the pulsar is unlikely. Thus, we conclude that in the sample of 14 pulsars only 2 pulsars, J0745--5353 and J1125--5825,  are likely detected.

\begin{table*}[]
\caption{Properties of the ordinary and millisecond pulsars that are investigated in this study.}
\label{tab:archive1}
\setlength{\tabcolsep}{12pt}
	\begin{center}
	\scalebox{0.65}{
	\begin{tabular}{cc ccc  ccc ccccc} \hline \T \B
		\multirow{2}{*}{Pulsar}  & \multirow{2}{*}{ObsID}  & $t_{\rm PN}$ & $t_{\rm M1}$ & $t_{\rm M2}$ & $b_{\rm PN}$ & $b_{\rm M1}$ & $b_{\rm M2}$ & {\it P} & {\it D}$_1$ & {\it D}$_2$ & log $\dot{E}$ & \multirow{2}{*}{{\it N}$_{\rm H, 21}$} \T \B \\
	&  & (ks) & (ks) & (ks) & (c/s) & (c/s) & (c/s) & (ms) & {\sml (kpc)} & {\sml (kpc)} & {\sml ({\it erg~s}$^{-1}$)} & \\
	
		\hline \T \B
		J0340$+$4130 & 0605470101 & 19.5 (6.3) & 20.7 (14.8) & 20.6 (15.0) & 0.4 (TN) & 0.35 (TN) & 0.35 (M) & 3 & 1.60$^*$ & 1.73 &  33.9 & 1.5   \\
		 \T \B
		J1435$-$5954 & 0692050101 & 130.1 (90.0) & $--$ & $--$ & 0.4 (M) & $--$ & $--$ & 473 & 1.06 & 1.18 & 32.7 & 1.3 \\
		 \T \B
		J1535$-$4114 & 0652610201 & 101.4 (63.2) & 102.4 (84.6) & 102.4 (83.9) & 0.4 (TK) & 0.35 (TK) & 0.35 (TK) & 432 & 2.77 & 1.95 & 33.3 & 1.99   \\
		  \T \B
		J1622$-$0315 $^B$ & 0784770401 &  19.0 (4.8) & 20.6 (16.5) & 20.6 (16.1) & 0.4 (M) & 0.30 (M) & 0.35 (M) & 4 & 1.14 & 1.11 & 33.9 & 0.6   \\
		 \T \B
		J1643$-$1224 $^B$ & 0742520101 & 21.8 (9.3) & 23.4 (13.8) & 23.4 (11.1) & 0.4 (M) & 0.35 (M) & 0.35 (M) & 5 & 0.74$^*$ & 2.40 & 33.8 & 1.87   \\
		  \T \B
		J1831$-$0952 & 0822330101 & $--$ & 66.6 (6.2) & 66.6 (8.7) & $--$ & 0.35 (M) & 0.35 (M) & 67 & 3.68 & 4.05 & 36.0 & 7.41   \\
		 \T \B
		J1857$+$0943 $^B$ & 0742520201 & 10.0 (5.3) & 11.6 (6.3) & 11.6 (8.8) & 0.4 (M) & 0.25 (M) & 0.25 (M) & 5 & 1.20$^*$ & 1.17 & 33.6 & 0.40  \\
		 \T \B
		J1926$-$1314 & 0742620101 & 75.0 (56.6) & 76.6 (54.4) & 76.6 (74.7) & 0.4 (TN) & 0.35 (TN) & 0.35 (TN) & 4864 & 1.53 & 1.48 & 31.1 & 1.22  \\
		\hline
	\end{tabular}
	}
	\end{center}
	\tablefoot{
	Pulsar names and their \emph{XMM-Newton} obsid's are given in the first column. The pulsars in binary systems are marked with $^B$. EPIC-pn and EPIC-MOS total and GTI-filtered exposure times (given in parentheses) are displayed in second column. 
The GTI filter for the background light curve and the corresponding filters are listed in third column for each EPIC detector. ``TN'' is thin filter, ``TK'' is thick filter and ``M'' is medium filter. Source properties are noted in the last column. $P$ is the spin period in seconds, {\it D}$_1$ is the best estimate of the pulsar distance in kpc according to the ATNF pulsar catalog which uses the \emph{YMW+17} DM-based distance as default, log $\dot{E}$ is the common logarithm of the spin-down energy loss rate in ergs$^{-1}$ and $N_{\rm H}$ is the hydrogen column density in unit of $ 10^{21}$ cm$^{-2}$ estimated from  the DM as outlined in \S\ref{ss:sa}. Sources with parallactic distance are marked with $^*$.}
\end{table*}

\section{Additional archival X-ray counterpart candidates of pulsars}\label{sec:hiddenp}

While using archival X-ray source catalogs to optimize our detection energy range, we noticed some potential pulsar counterparts which to our best knowledge, have not been reported. To identify them, we first matched the radio position of pulsars in the latest ATNF catalog (Manchester et al. \citeyear{Manchester2005AJ}, V1.64) with the \emph{XMM-Newton} \citep[4XMM DR10,][]{Webb2020} and \emph{Chandra} \citep[CSC v2,][]{Evans2010ApJS} catalogs, and selected all the pulsars that fall in the search radius of 10$\arcsec$ and 5$\arcsec$ of catalogs respectively. We selected 340 X-ray sources. Assuming an upper limit of 2 kpc on pulsar distance,  the number of candidates is reduced to 101. Finally, we excluded all pulsars with known X-ray counterparts using archives such as SIMBAD and ADS. As a result, we identified new potential X-ray counterparts for 3 nearby pulsars in the 4XMM-DR10 catalog not reported in the literature.

Furthermore, we used the same catalogs to also look for new significant detections of pulsars for which upper limits were reported in X-ray surveys such as carried out by \cite{Prinz2015} and  \cite{Lee2018ApJ}.  
Typically, such new detections are due to additional data, or improved SAS/CIAO processing routines. We found 5 additional \emph{XMM-Newton} catalog sources within a 10$\arcsec$ search radius of the radio pulsar positions). We did not find  any unpublished X-ray emitting pulsar candidate in CSC v2. For consistency, for all the archival sources, we performed the same analyses as we did for our survey pulsars. This allowed a direct comparison in terms of energy bands and fluxes. The source properties and analyses results are provided in Table \ref{tab:archive1} and Table \ref{tab:archive2}. The fluxes are given in 0.3-2.0 keV for BB model and 1-10 keV for PL model in Table \ref{tab:source_list3}. 
The total positional uncertainty of archival sources are obtained from the 4XMM-DR10 catalog. We use the \texttt{POSERR} at the $3\sigma$ confidence level. We note that two of our archival sources have a distance > 2 kpc based on \emph{YMW+17} model but we decided to include them due to their interesting X-ray characteristics.

\begin{table*}[]
\caption{Properties of ordinary and ms-pulsars with possible counterparts in 4XMM-DR10 catalog for which we could not find a respective note in the literature or previously marked as upper-limit.}
\label{tab:archive2}
\setlength{\tabcolsep}{12pt}
	\begin{center}
	\scalebox{0.67}{
	\begin{tabular}{c c c c cc cc cc cc} \hline \T \B
	\multirow{3}{*}{\large{Pulsar}} & \multirow{3}{*}{\large{ObsID}} & 
	\multirow{2}{*}{\large{\textbf{$\Delta_{\rm X-R}$}}} & \multirow{2}{*}{\large{$\bm{\theta}$}} &
	\multirow{3}{*}{\large{IAU name}} &
	\multirow{3}{*}{\large{ML}} &
	\multicolumn{2}{c}{PN$_{\rm \,0.3-2\,keV}$} &
	\multicolumn{2}{c}{MOS1$_{\rm \,0.3-2\,keV}$} &
	\multicolumn{2}{c}{MOS2$_{\rm \,0.3-2\,keV}$} \T \B  \\
	 & & & & & & $N_{\rm s}$ & $S$ & $N_{\rm s}$ & $S$ & $N_{\rm s}$ & $S$ \T \B \\
	 & & (arcsec) & (arcmin) & & & (counts) & ($\sigma$) & (counts) & ($\sigma$) & (counts) & ($\sigma$)  \T \B \\
		\hline \T \B
		J0340$+$4130 & 0605470101 & 4.74 & 7.8 & 4XMM J034023.3+413040 & 12.0 &
        11.4$\pm$5.4 & 2.1 &
        13.9$\pm$5.6 & 2.5 &
        13.6$\pm$4.3 & 3.1  \\
	    \T \B
		J1435$-$5954 & 0692050101 &  2.59 & 16.9 & 4XMM J143500.1-595452 & 11.3 &
        66.8$\pm$17.9 & 3.7 &
        $--$ & $--$ &
        $--$ & $--$ \\
	    \T \B
		J1622$-$0315 & 0784770401 &  0.84 & 1.7 & 4XMM J162259.6-031538 & 63.5 & 
        33.9$\pm$7.8 & 4.3 &
        24.1$\pm$6.8 & 3.5 &
        20.1$\pm$6.9 & 2.9 \\
		 \T \B
		J1535$-$4114 & 0652610201 &  5.6 & 4.3 & 4XMM J153516.5-411402 & 12.7 & 
        72.5$\pm$16.5 & 4.4 &
        21.0$\pm$9.8 & 2.1 &
        19.8$\pm$8.5 & 2.3 \\
		 \T \B
		J1643$-$1224 & 0742520101 & 1.5 & 1.1 & 4XMM J164338.0-122458 & 112.7 & 
        66.8$\pm$11.3 & 5.9 & 
        24.4$\pm$6.4 & 3.8 &
        22.3$\pm$6.3 & 3.5 \\
		 \T \B
		J1831$-$0952 & 0822330101 & 2.2 & 2.3 & 4XMM J183134.1-095201 & 96.9 &
        $--$ & $--$ & 
        28.9$\pm$9.6 & 3.0 & 
        29.0$\pm$9.7 & 3.0 \\
		 \T \B
		J1857$+$0943 & 0742520201 & 1.3 & 1.2 & 4XMM J185736.4+094317 & 38.3 &
        27.9$\pm$6.8 & 4.1 & 
        7.1$\pm$4.5 & 1.6 &
        13.8$\pm$4.7 & 2.9 \\
		 \T \B
		J1926$-$1314 & 0742620101 & 5.5 & 1.1 & 4XMM J192653.7-131358 & 6.9 &
        59.6$\pm$16.7 & 3.5 & 
        13.5$\pm$8.2 & 1.6 & 
        27.6$\pm$9.6 & 2.8 \\
        \hline
	\end{tabular}
	}
	\end{center}
	\tablefoot{$\Delta_{\rm X-R}$ column displays angular separation in $\arcsec$ between the pulsar's radio timing position and the centroid of the nearest X-ray source according to 4XMM-DR10. $\bm{\theta}$ is the EPIC-pn off-axis angle. ML is the reported 4XMM-DR10 maximum likelihood in the energy band 0.2-12.0 keV. Background-corrected counts $N_{\rm s}$ at the optimized X-ray source position and their corresponding significance are displayed for EPIC-pn and EPIC-MOS1/MOS2. PSR J1831--0952 has a $\sim$ $6\sigma$ significance in 0.3-4.5 keV with Epic-MOS cameras. The off-axis angle is given for Epic-MOS1.}
\end{table*}

   \begin{figure*}
   \centering
   \includegraphics[width=0.24\textwidth]{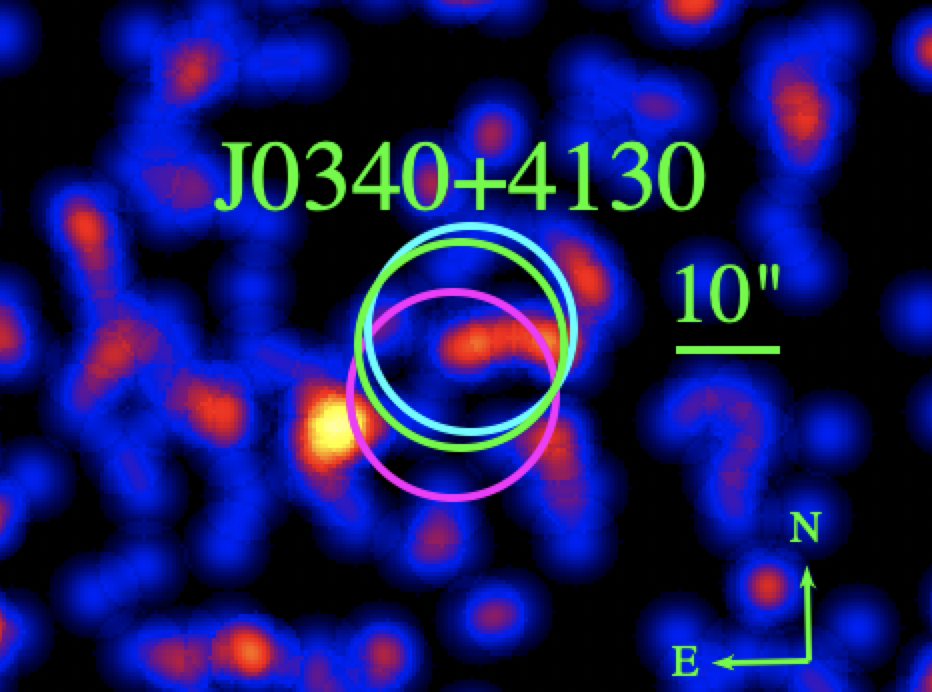}
   \includegraphics[width=0.24\textwidth]{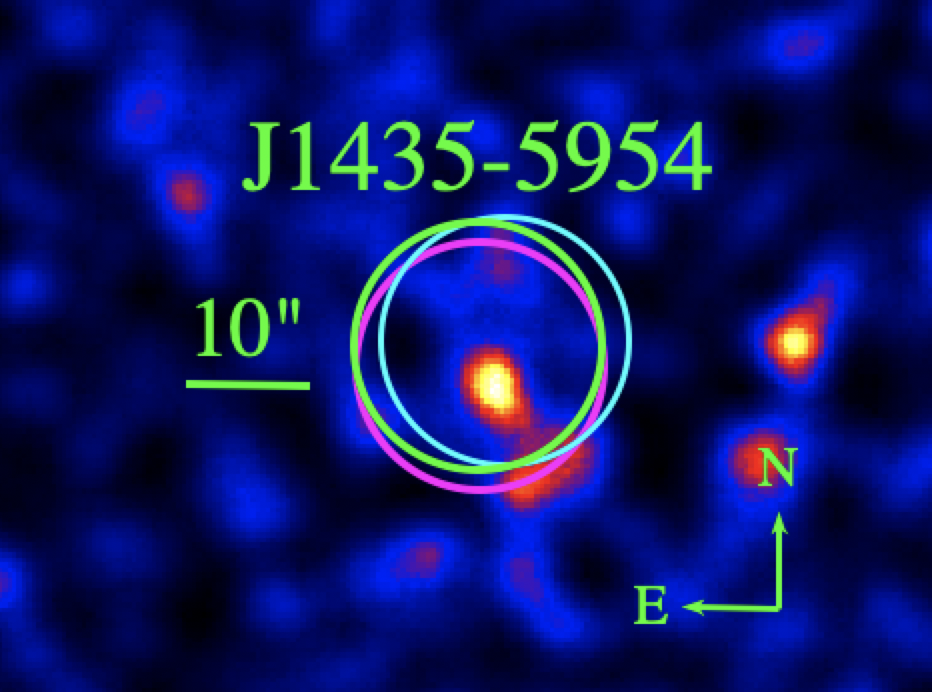}
   \includegraphics[width=0.24\textwidth]{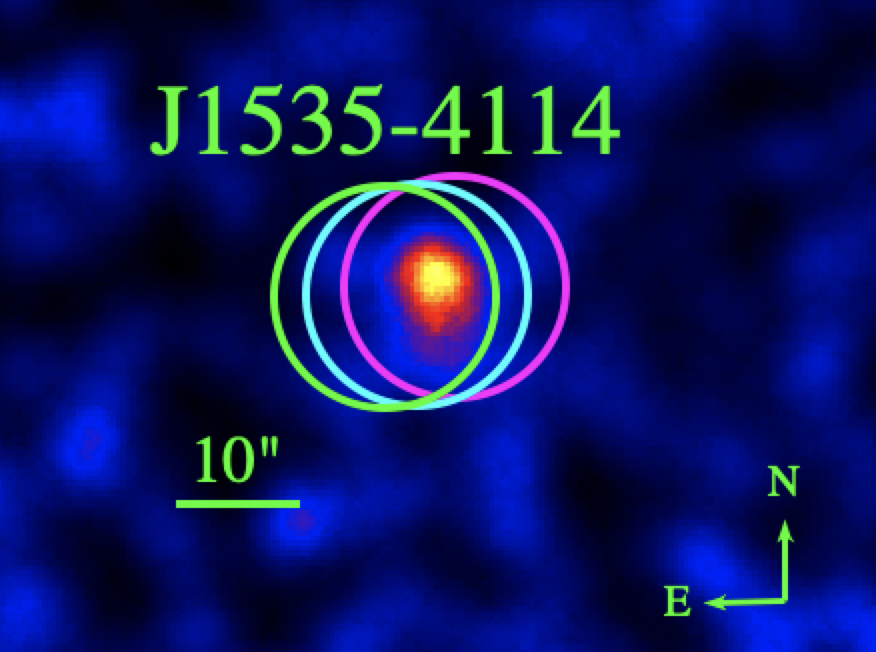}
   \includegraphics[width=0.24\textwidth]{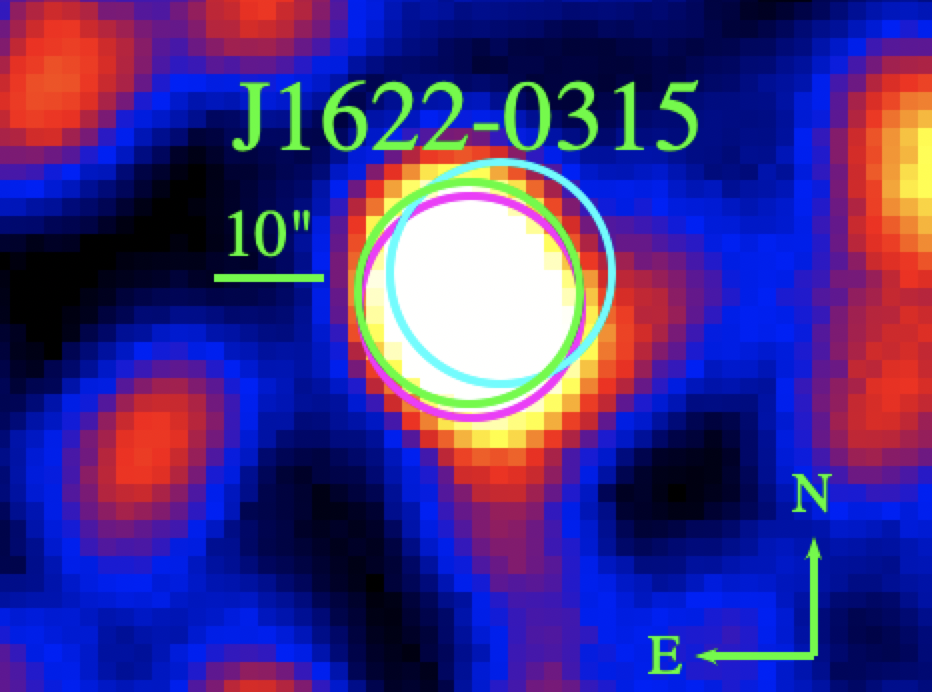}\\
   \includegraphics[width=0.24\textwidth]{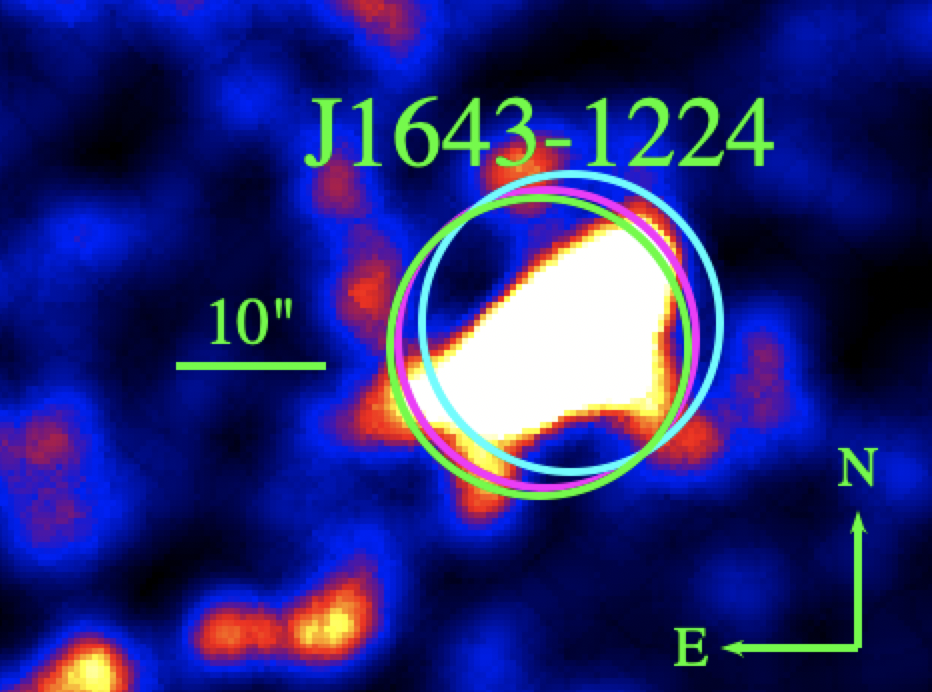}
   \includegraphics[width=0.24\textwidth]{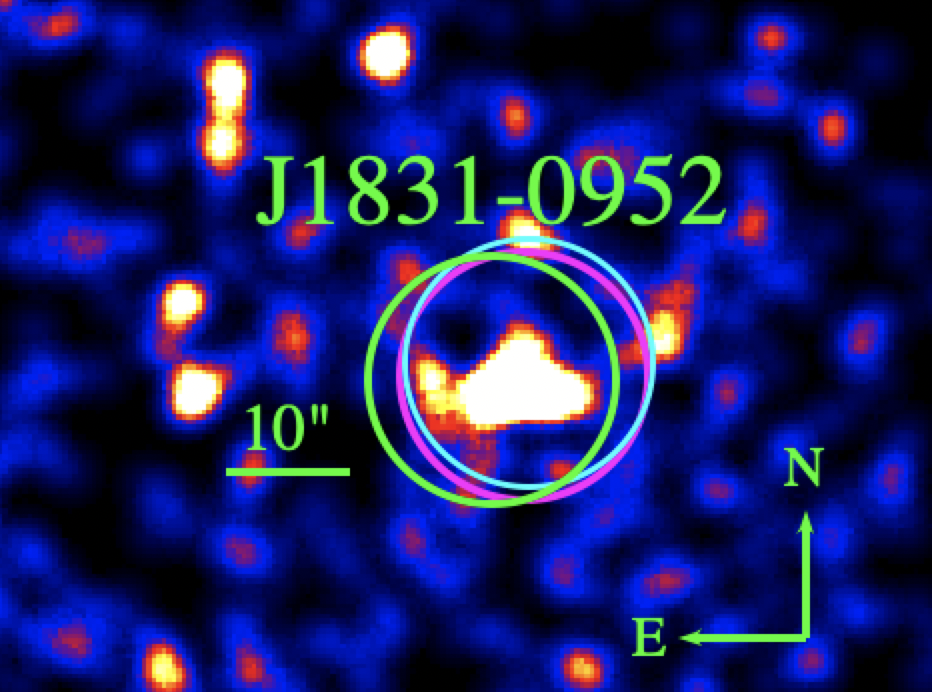}
   \includegraphics[width=0.24\textwidth]{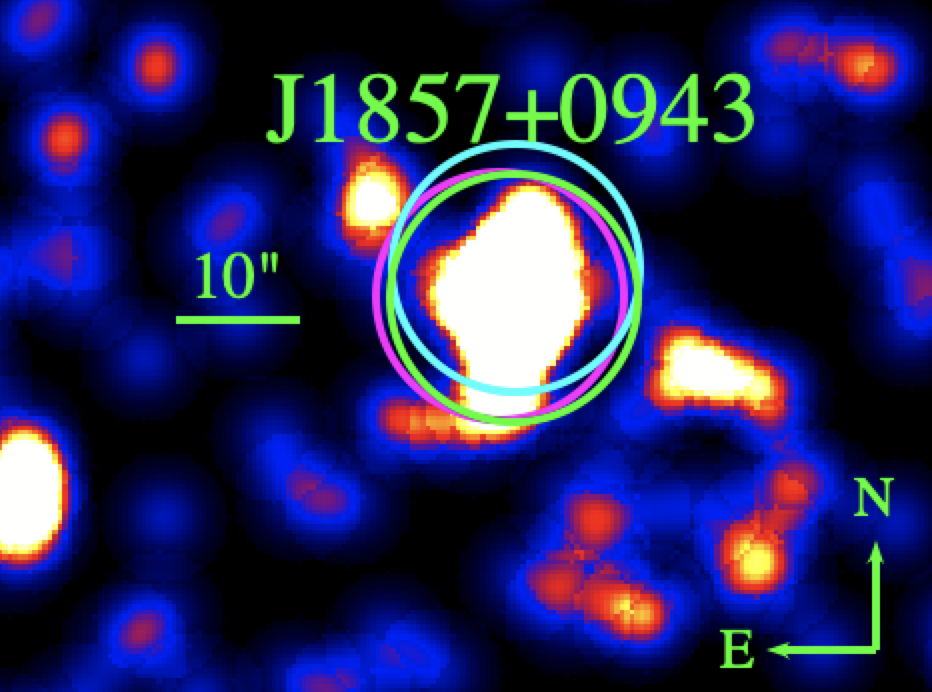}
   \includegraphics[width=0.24\textwidth]{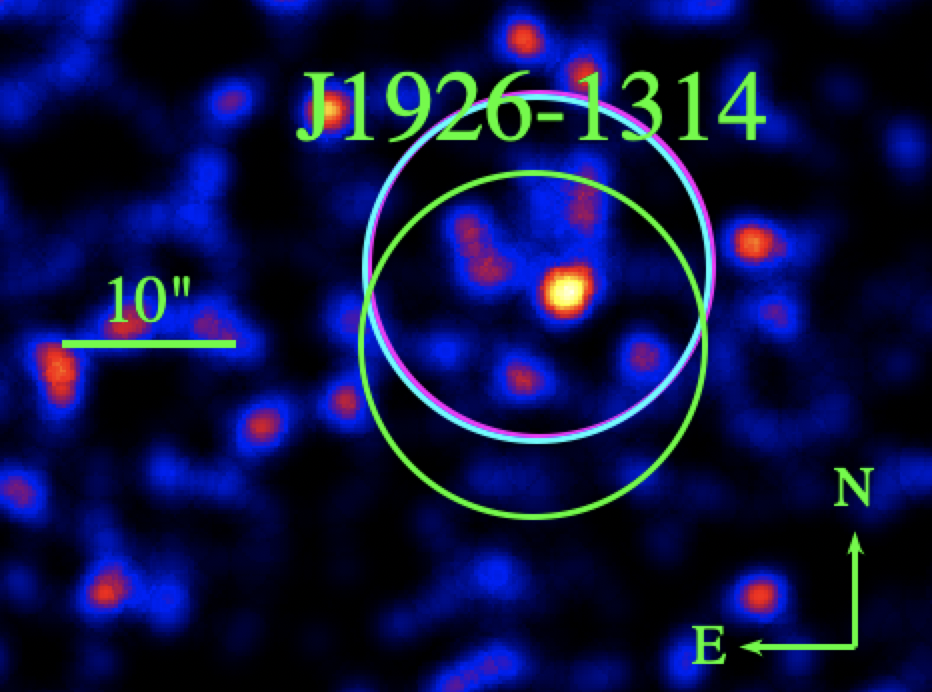}
   \caption{X-ray counterpart candidates in 0.3-2.0 keV as seen with the EPIC-pn (except for J1831 and J1535 which are displayed in EPIC-MOS1).
            The green circles are centered at the radio timing position of the pulsars whereas the blue circles are centered at the X-ray source with the highest significance. The Magenta circles show the 4XMM-DR10 positions close to these pulsars. The images have pixel scales of $0.4\arcsec$\, and {are} smoothed with Gaussian length scales $\sigma$ =2.2$\arcsec$
            (except PSRs\,J1535--4115 and J1926--1314 with $\sigma$ =1.4$\arcsec$). The angular separations between X-ray and radio positions are given in Table \ref{tab:archive2}.}
              \label{fig:archive}%
    \end{figure*}

\begin{table*}[]
\caption{X-ray fluxes and luminosities of the investigated sources.}
\label{tab:source_list3}
\setlength{\tabcolsep}{12pt}
	\begin{center}
	\scalebox{0.95}{
	\begin{tabular}{c cc cc} \hline \T \B
	\multirow{3}{*}{Pulsar}   & 
    \multicolumn{2}{c}{Blackbody  ({\it kT} = 0.3\,keV)} &
	\multicolumn{2}{c}{Power-law ($\Gamma = 1.7$)}  \T \B \\
	& $F^{\rm unabs}_{\rm{0.3-2\,keV}}$ &
	$L^{\rm{BB}}_{\rm{0.3-2\,keV}}$ &
	$F^{\rm unabs}_{\rm{1-10\,keV}}$ &
	$L^{\rm{PL}}_{\rm{1-10\,keV}}$ \T \B \\
 & ({\sml $10^{-14}$ erg s$^{-1}$ cm$^{-2}$}) & ({\sml $10^{29}$ erg s$^{-1}$}) & ({\sml $10^{-14}$ erg s$^{-1}$ cm$^{-2}$}) & 
	({\sml $10^{29}$ erg s$^{-1}$}) \T \B\\
		\hline \T \B
		J0340$+$4130 & 0.46$\pm$0.23 & 14.1$\pm$7.1  & 1.07$\pm$0.58  & 33.4$\pm$18.1\\
		  
		\T \B
		J0711--6830 & 0.58$\pm$0.32 & 0.10$\pm$0.06 & 2.75$\pm$1.22 & 0.44$\pm$0.21 \\
		\T \B
		J0745--5353 & 0.80$\pm$0.23 & 3.15$\pm$0.90 & 0.92$\pm$0.33 & 3.58$\pm$1.21  \\
		\T \B
		J0942--5552 & 1.38$\pm$0.43 & 1.48$\pm$0.52 & 0.25$\pm$0.30 & 0.23$\pm$0.40\\
		\T \B
		J1125--5825 & 1.36$\pm$0.48 & 49.3$\pm$17.5 & 2.45$\pm$1.44 & 88.9$\pm$51.2 \\
		\T \B
		J1435--5954 & 0.45$\pm$0.14 & 6.05$\pm$1.3 & 0.59$\pm$0.23 & 7.93$\pm$3.12  \\
		\T \B
		J1535--4114 & 0.31$\pm$0.10 & 28.5$\pm$6.4 & 0.34$\pm$0.10 & 31.2$\pm$10.1 \\
		\T \B
		J1622--0315 & 0.85$\pm$0.24 & 13.2$\pm$2.9 & 1.30$\pm$0.42 & 20.2$\pm$6.5\\
		\T \B
		J1643--1224 & 1.58$\pm$0.27 & 10.4$\pm$1.8 & 2.23$\pm$0.51 & 14.6$\pm$3.2 \\
		\T \B
		J1831--0952 & 0.41$\pm$0.14 & 66.4$\pm$22.7 & 2.36$\pm$0.40 & 371$\pm$60 \\
	    \T \B
		J1857$+$0943 & 0.84$\pm$0.23 & 13.8$\pm$3.4 & 0.54$\pm$0.22 & 9.4$\pm$4.9 \\
		\T \B
		J1926--1314 & 0.20$\pm$0.05 & 5.8$\pm$1.4 & 0.15$\pm$0.11 & 4.3$\pm$3.1  \\
			  
	    \T \B
		 J0942--5552 & $<$1.21 & $<$1.30 & $<$2.69 & $<$2.90 \\
		\T \B
		 J0945--4833 & $<$0.79 & $<$1.16 & $<$1.74 & $<$2.55 \\
		\T \B
		 J0954--5430 & $<$1.22 & $<$2.70 & $<$2.65 & $<$5.86 \\
		\T \B
		 J0957--5432 & $<$1.57 & $<$3.80 & $<$3.38 & $<$8.19 \\
		\T \B
		J1000--5149 & $<$0.19 & $<$0.03 & $<$0.42 & $<$0.08 \\
		\T \B
		 J1003--4747 & $<$0.63 & $<$1.03 & $<$1.39 & $<$2.28 \\
		\T \B
		 J1017--7156 & $<$0.49 & $<$0.40 & $<$1.07 & $<$0.87 \\
		\T \B
		 J1543--5149 & $<$0.42 & $<$6.65 & $<$0.90 & $<$14.2 \\
		\T \B
		 J1725--0732 & $<$0.15 & $<$0.07 & $<$0.32 & $<$0.15 \\
		\T \B
		 J1740--3015 & $<$2.96 & $<$5.67 & $<$6.50 & $<$12.4 \\
		\T \B
		 J1755--0903 & $<$0.81 & $<$0.51 & $<$1.75 & $<$1.11 \\
		 \hline
	\end{tabular}
	}
	\end{center}
	\tablefoot{The flux values are based on the EPIC-pn count rates (or limits) and two spectral model assumptions. $F^{\rm unabs}$ represents the unabsorbed X-ray flux whereas $L^{\rm{BB}}$ and $L^{\rm{PL}}$ columns represent the X-ray BB and PL luminosities respectively. For the luminosity calculation, distances based on the \cite{Yao2017ApJ} electron density model are used. The distance uncertainty is not taken into account for error estimation. For PSR J1831--0952, spectral parameters have been obtained (see Table~\ref{tab:1831}) but here we list the flux derived from typical model parameters for consistency.}
\end{table*}

\subsection*{PSR J1622--0315}

The redback millisecond pulsar was observed in 2017 in a 22\,ks exposure with EPIC and reported to be fitted well with a PL ($\Gamma$ = 2.0$\pm$0.3) in the PhD thesis by \citet{Gentile2018PhDT}. The X-ray counterpart candidate is located $0.8\arcsec$ away from the radio pulsar position. 

Considering a total position uncertainty of 3.3$\arcsec$ ($3\sigma$), we found one \emph{Gaia} source $0.9\arcsec$ away from the X-ray source which can be excluded as the sole  counterpart to the X-ray source (Table \ref{gaia}). The Gaia source is reported as the companion counterpart of the pulsar \citep{Strader2019ApJ}, but the X-ray emission is  expected to come from the pulsar and possibly from an intra-binary shock (e.g., \citealt{Romani2016}). We can also exclude
this source as the (main) X-ray counterpart of the pulsar based on the contradiction between the expected temperature of an M star  ($\sim$ 3200 k \citealt{Rajpurohit2013AA}) and the reported temperature  $T_{eff} =6108 K$ according to \citealt{Stassun2019AJ}. There is also one $\rm{WISE}$ source (ID: J162259.63--031536.9) with an angular separation of $1.2\arcsec$ and due to the lack of $\rm{WISE}$ colors, we are not able to classify the nature of this source.

   \begin{figure}
   \centering
    \includegraphics[width=0.49\textwidth]{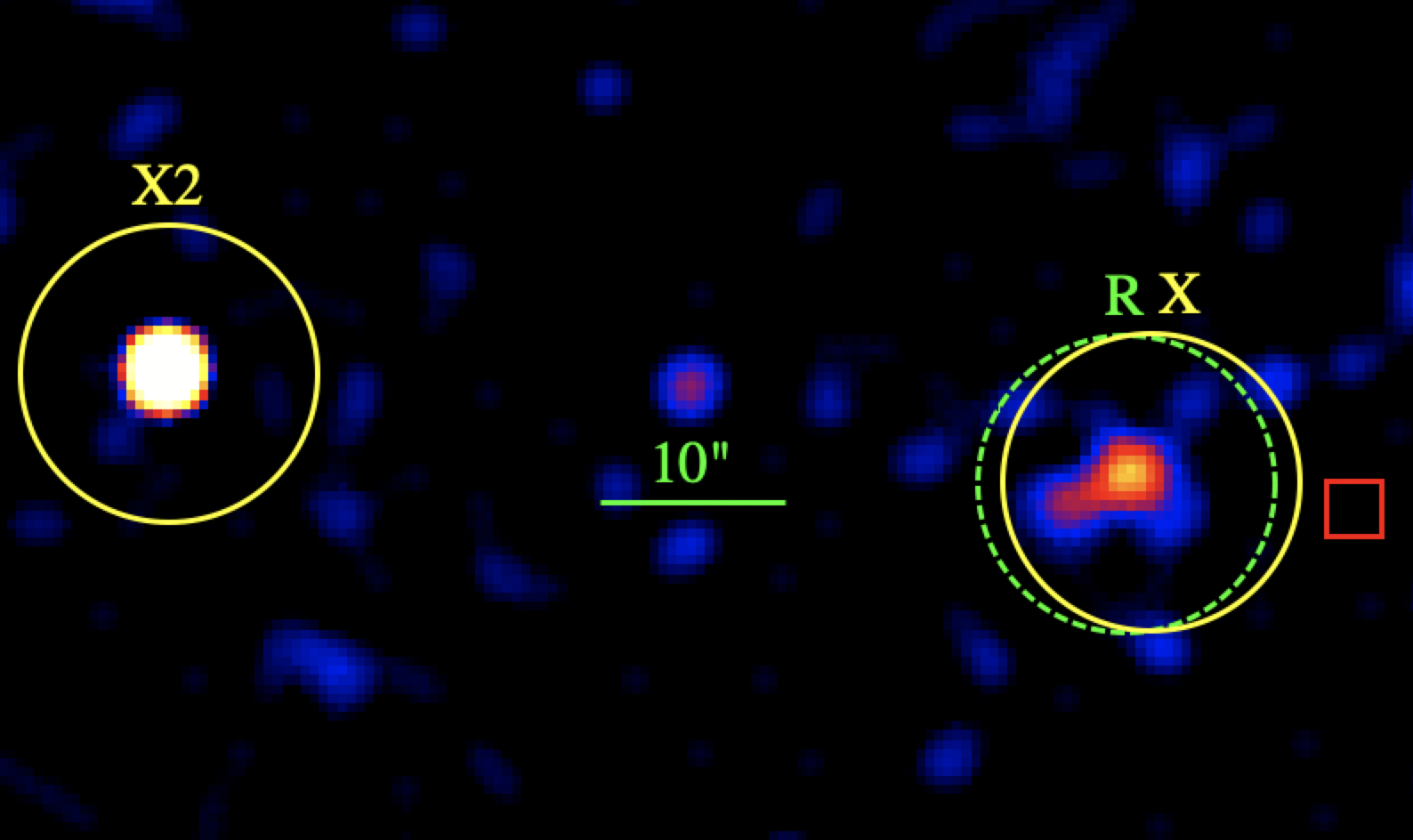}
      \caption{\emph{Chandra} image of PSR J1831--0952 and X2, a point source in the field of view. 
      The green circle is centered at the radio timing position of the pulsar whereas the yellow circles are centered at the positions of \emph{Chandra} X-ray sources (with off-axis angle of 12.2$\arcsec$ for the extended source, X, and $\ang{;1.1;}$ for X2) with the highest significance in \emph{Chandra} observations. The image has a pixel scale of 0.49$\arcsec$ and smoothed with a Gaussian length scale of $\sigma$ =1.25$\arcsec$. The red square shows the position of the \emph{Chandra} optical axis. Despite the X-ray counterpart of the pulsar being much closer to the optical axis, only X2 has a point-like shape, not showing a ``tail'' as the pulsar.}
         \label{fig:J1831_chandra}
   \end{figure}

\subsection*{PSR J1831--0952}
\label{ssec:j1831}

Since the X-ray counterpart candidate has 155$\pm$40 combined net counts in the energy range  0.3-10\,keV, we performed a spectral fit. \emph{XMM-Newton} only covered the source with EPIC-MOS. For the fit we combined EPIC-MOS1 and EPIC-MOS2 spectra in the 0.3-10\,keV energy range. We fitted the spectra with a single PL by first fixing $N_{\text{H}}$ to DM-derived 7.4 $\times$ 10$^{21}$ atoms cm$^{-2}$ and then freeing it. In both cases, the photon index is < 1.3, consistent with emission from a pulsar (Table \ref{tab:1831}). Since $N_{\text{H}}$ value is unconstrained, we only list the latter case. 
We also carried out BB-fits. The temperature we obtained from the fit is $\sim$ 1.2 keV which is relatively high. Using the spectral fit, we estimated an unabsorbed flux $f$, $f_{-14}= f / 10^{-14}$ erg s$^{-1}$ = 4.3$\pm$0.4 in 1-10 keV  for the PL and  $N_{\text{H}}$ from Table \ref{tab:archive1}. The reported errors are estimated for the 90\% confidence level using the steppar command. \cite{Abichandani2019ATel} briefly reported
the \emph{Chandra} detection of a possibly extended source which is coincident with the \emph{XMM-Newton} source (XGPS-I J183134--095155). We also investigated the corresponding \emph{Chandra} observation. The source has been observed in VFAINT mode with ACIS-I for 29 ks. We obtained 30.6 net counts in 0.3-8.0 keV. The \emph{Chandra} image of the source is shown in Figure \ref{fig:J1831_chandra}. The image indeed indicates that the source is more extended than a typical point source with the extension of  $\sim 10\arcsec$. Most of these counts ($\sim $60\%) of this extended emission are detected in the energy range 2.4-8.0 keV. Considering this is likely an extended emission, the PWN could be the main contributor of the non-thermal emission.

\begin{table}[]
\caption{Spectral fit parameters for PSR J1831--0952 obtained from combined MOS data.}
\label{tab:1831}
\setlength{\tabcolsep}{12pt}
	\begin{center}
	\scalebox{0.75}{
	\begin{tabular}{cccccc} \hline \T \B
		\multirow{2}{*}{Model}  & $\Gamma$ / kT &  
	\multirow{2}{*}{N$_{\text{H,21}}$} & \multirow{2}{*}{F$_{-14}^{\text{unabs}}$} & \multirow{2}{*}{F$_{-14}^{\text{abs}}$} & \multirow{2}{*}{$\chi^2$/d.o.f} \T \B \\
	& / (keV) & &  &  & \\
	\hline \T \B
		PL & 0.9$\pm$0.3 & 7.4 & 4.3$\pm$0.4 & 4.0$\pm$0.7 & 17.8/15 \\
	\T \B
		BB & 1.2$\pm$0.2 & 7.4 & 0.4$\pm$0.1 & 0.2$\pm$0.04 & 12.3/15\\
		\hline
	\end{tabular}
	}
	\end{center}
	\tablefoot{
	Absorbed and unabsorbed fluxed are calculated in 1-10 keV for PL model and in 0.3-2 keV for BB model. Errors are estimated with a 90\% confidence interval.}
\end{table}

\subsection*{PSR J1926$-$1314}
\label{ssJ1926}

The solitary ({$\tau=2$}\,Myr) pulsar J1926--1314 was discovered by \citet{Rosen2013ApJ} with a relatively high inferred magnetic dipole field of $B_{\rm surf}=1.4 \times 10^{13}$\,G. A faint, soft X-ray source, is detected $5.4\arcsec$ from the pulsar.

If this X-ray source is indeed the counterpart of the pulsar J1926--1314, and its DM-distance estimate (1.5\,kpc) is correct, then the estimated X-ray efficiency (above 0.01 for the chosen spectral model parameters, see Figure~\ref{fig:xrayeff1}) is unusually high. Considering the inferred magnetic dipole strength of $10^{13}$\,G, such a result is interesting with respect to the suggested hypothesis that the magnetic field affects the observed thermal properties of pulsars (e.g., \citealt{Olausen2013, Pons2009}).

However, there are three \emph{Gaia} $\rm{TESS}$ sources within $5\arcsec$ of the X-ray source (Table~\ref{gaia}), and we cannot exclude that any of them is the actual (stellar) counterpart of the X-ray source or contributes to the X-ray flux. Detecting pulsations of the X-ray source would confirm the pulsar counterpart. For this, however, more X-ray counts, i.e., a longer observation would be required.

\begin{table*}
    \caption{ \emph{Gaia}-$\rm{EDR3}$ sources
    that  fall within $3\sigma$ positional uncertainty radius of the X-ray sources.}
	\begin{center}
	\scalebox{0.68}{
	\begin{tabular}{c c c c c c c c c c c c c}
	 \hline \T \B
		\emph{Gaia}-$\rm{EDR3}$ & Associated PSR & $ \mu_{\alpha}$ & $ \mu_{\delta}$ & $D$ & $T_{\rm eff}$ & $V$ & log ($f_{\rm x} /f_{\rm v}$) & Macc88 & CV07 & BJ19 & \multicolumn{2}{c}{Excluded}\T \B \\
		 & & (mas/y) & (mas/y) & (kpc) & (K) & (mag) & & &  \small{(Spectral type)}  & & (S) & (GXY) \T \B \\
		\hline \T \B
    237123598225544064 & J0340+4130 (4.64$\arcsec$) & 
		-0.5$\pm$0.4 & 
		-2.6$\pm$0.4 & 2.0$\pm$1.1 & 4154$\pm$182 & 20.6$\pm$0.7 &
		 -0.5 & AGN &
		$--$ &
		S, QSR & Y & Y\\
			 \T \B
    5339616527672093056 & J1125--5825 (1.27$\arcsec$) & 
		-4.8$\pm$0.1 & 
		1.1$\pm$0.1 & 4.0$\pm$2.0 & $--$ & 18.7$\pm$0.4 &
		 -0.85 & M, AGN, GXY &
		M1, M5, M2 &
		S, QSR & N & Y\\
		 \T \B
		 5339616527704864768 & J1125--5825 (3.85$\arcsec$) & 
		-6.8$\pm$0.1 & 
		1.0$\pm$0.1 & 1.7$\pm$0.1 & 5488$\pm$393 & 16.4$\pm$0.3 & -1.7 &
		 M, K, GXY &
		G8 & S, GXY & N & Y\\
		 \rowcolor{gray!1} \T \B
    5878785517750906240 & J1435--5954 (4.35$\arcsec$) & 
		-9.5$\pm$0.1 & 
		-2.4$\pm$0.1 & 2.2$\pm$1.1 &  6635$\pm$180 & 19.1$\pm$0.4 &
		 -1.3 & M, GXY &
		M6, M7, M8 &
		S, GXY & N & Y\\
		 \T \B
		 5878785517754291968 & J1435--5954 (6.83$\arcsec$) & 
		-6.5$\pm$0.2 & 
		-2.6$\pm$0.1 & 2.3$\pm$1.1 &  5456$\pm$212 & 19.9$\pm$0.5 & -0.9 &
		 M, AGN, GXY &
		M5, M6, K7 &
		S, GXY, QSR & N & Y\\
		 \T \B
   4358428942492430336 & J1622--0315 (0.89$\arcsec$) & 
		-13.2$\pm$0.3 & 
		2.3$\pm$0.2 & 2.7$\pm$1.9 &  6108$\pm$254 & 19.6$\pm$0.4 &
		 -0.7 & M, GXY, AGN &
		$---$ &
		S & Y & Y\\
		 \T \B
    4310888159960965632 & J1857$+$0943 (2.71$\arcsec$) & 
		-1.9$\pm$0.6 & 
		-4.1$\pm$0.6 & 1.5$\pm$0.8  &  $--$ & 21.5$\pm$0.5 &
		 -0.2 & AGN &
		$---$ & S, GXY, QSR & Y & Y\\
		 \T \B
   4186460380401976448 & J1926--1314 (0.96$\arcsec$) & 
		$--$ & 
		$--$ & $--$ &  $--$ & 18.2$\pm$0.5 &
		 -2.1 & M,K &
		$--$ & S, QSR & N & Y\\
		 \T \B
    4186460380401976576 & J1926--1314 (1.96$\arcsec$) & 
		5.6$\pm$0.9 & 
		-4.9$\pm$0.5 & 4.2$\pm$2.3 &  $--$ & 14.7$\pm$0.4 &
		 -1.8 & M,K &
		$--$ &
		S, QSR & N & Y\\
		 \T \B
    4186460380410768640 & J1926--1314 (2.97$\arcsec$) & 
		2.4$\pm$0.1 & 
		-5.4$\pm$0.1 & 1.5$\pm$0.1 &  5861$\pm$123 & 18.4$\pm$0.5 &
		 -3.5 & B-F,G,K &
		M3, M4, F-8 & S, GXY & N & Y\\
		\hline
	\end{tabular}
	}
	\end{center}
	\tablefoot{
	The respective pulsars are listed in the second column, followed by the angular distance to the \emph{Gaia}-$\rm{EDR3}$ source in parentheses. The next columns show the proper motions, distances and effective temperatures obtained from $\rm{TESS}$ Input Catalog - v8.0 \citep{Stassun2019AJ}. The $V$-band magnitude (Vega) was estimated from the \emph{Gaia} magnitudes using Johnson-Cousins relationships of Landolt standard stars that were observed with \emph{Gaia} according to \cite{Evans2018A&A}. log ($f_{\rm x}/f_{\rm v})$ is the X-ray to optical flux ratio.  The Macc88, CV07 and BJ19 columns represent classifications according to \cite{Maccacaro1988ApJ} \cite{Covey2007AJ} and \cite{Jones2019} respectively. S, GXY, QSR and AGN represents star, Galaxy, Quasar and active galactic nucleus, respectively. For all investigated optical/NIR sources, \cite{Stassun2019AJ} reported a star classification in the TESS input catalog. \label{gaia}}
\end{table*}

\section{Discussion}
\label{sec:discussion}
The X-ray analyses of our radio pulsars revealed a few important aspects that explain  the low detection rate in the sample. First, more observation time than anticipated was lost due to background flaring events, leading to much shorter scientifically useful exposures. For instance, for 4 pulsars in our list (J0954--5430, J0957--5432, J1125--5825, and J1740--3015) we lost nearly 65\% of exposure times in the most sensitive detector.  

Second, our assumptions for estimating expected fluxes could be too optimistic. We assumed certain spectral models to compute the expected fluxes of ordinary and ms-pulsars. We also assumed an isotropic emission from the compact object for flux/luminosity conversion which may be substantially different in reality. There is another important, but rather uncertain parameter that has a substantial impact on the calculated fluxes -- the distance.

\subsection{Distance}
\label{ssec:distance}

Since none of our survey sources had a reliable parallax measurement, we used distances that are based on the DM and the \emph{YMW+17} electron density model.
\emph{YMW+17} noted that for high-latitude pulsars, the model benefited considerably from recent parallax measurements of VLBI and pulsar timing array projects, resulted in smaller distance errors compare to \emph{NE2001}-based distance. 

In our sample, for $\sim$ 70\% of sources, the distances differ by more than a factor of $\sim 2$ between the two models with the \emph{NE2001} model mostly resulting in the larger distance estimates. 
Interestingly, 57\% of our target list would not have satisfied the 2 kpc criteria if we have had used the \emph{NE2001} model for the distances. Among our detected sources, for J0745--5353 and J1125--5825, the distances provided by the two models differ by a factor of $\lesssim 2$ whereas for J0711--6830 the \emph{NE2001}-based distance is $\sim 8$ times higher.

Based on our detection rates, we conclude that we may have underestimated distances in our initial assessments. An improved approach would be to compare different estimates and only observe targets that either have parallax measurements, have small distance deviations (< factor 2) for different DM models, or are detectable for the largest listed distance.

\subsection{X-ray efficiency}
\label{ssec:discussx}
\begin{figure*}
   \centering
    \includegraphics[width=0.49\textwidth]{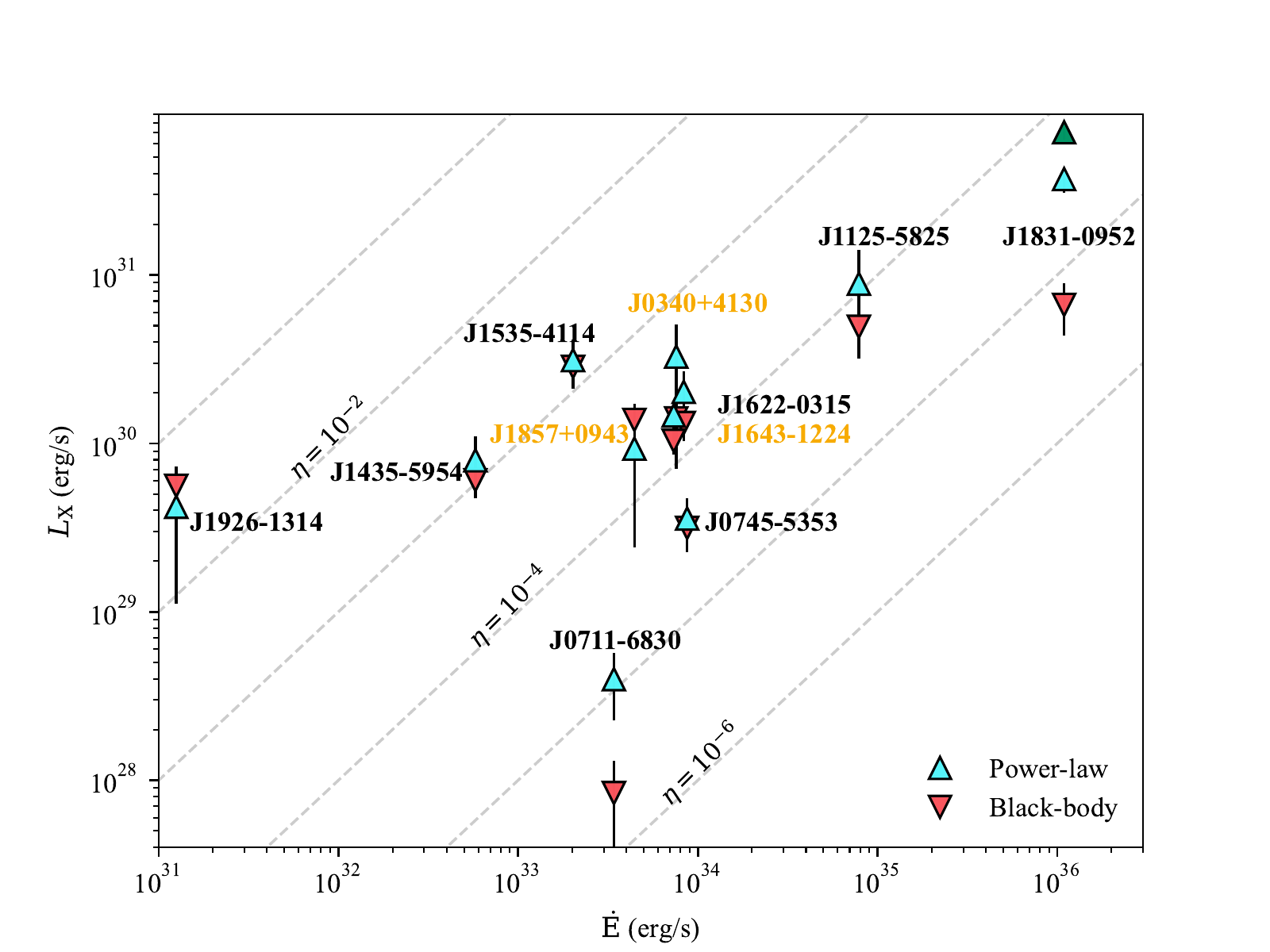}
    \includegraphics[width=0.49\textwidth]{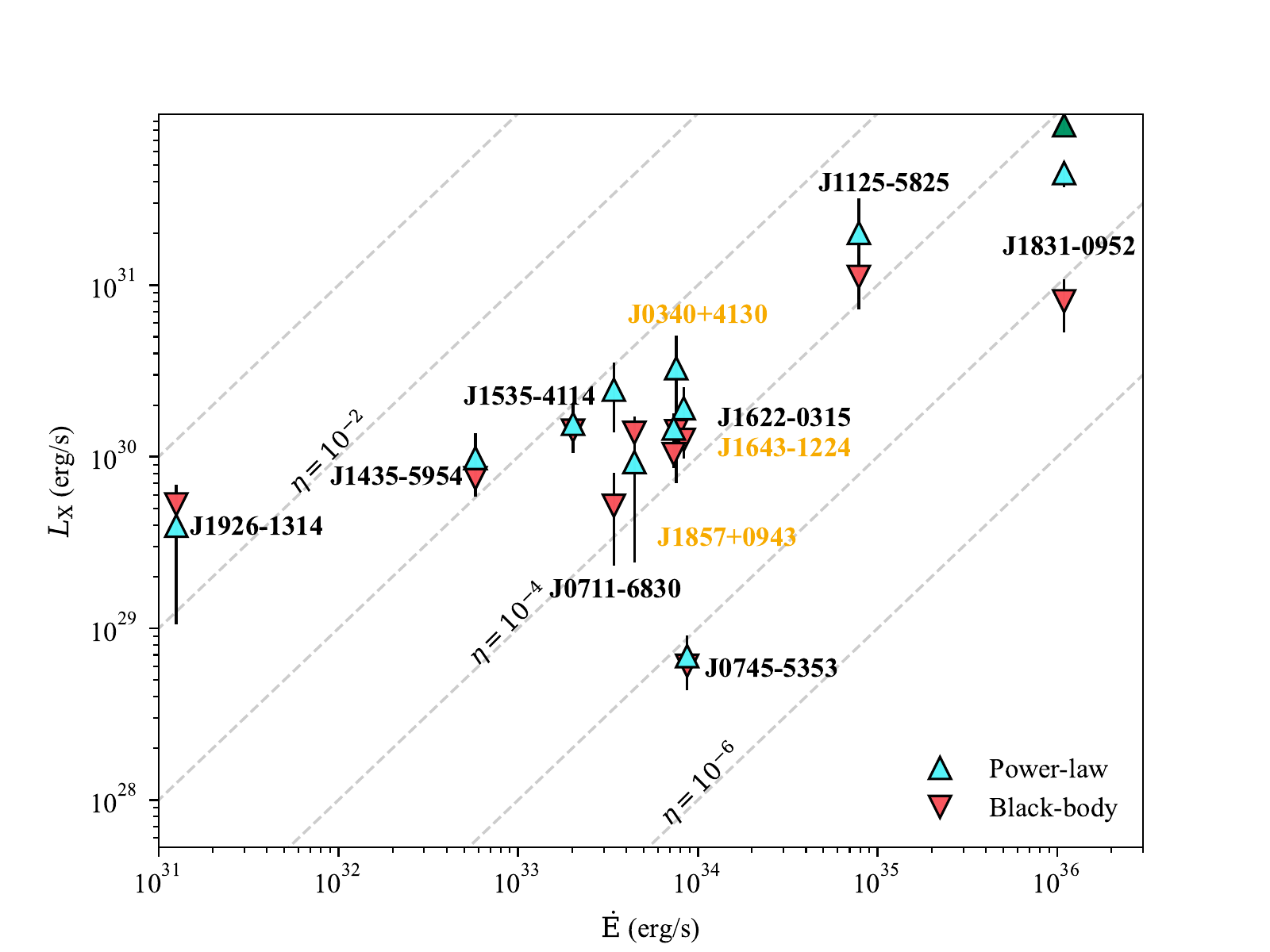}
      \caption{X-ray luminosity of ordinary and millisecond pulsars investigated in this study vs. their spindown powers $\dot{E}$ calculated with \emph{YMW+17} based distances (left) and \emph{NE2001} based distances (right).
      Sources with parallax measurements are displayed with yellow. The up-pointing triangles display the 1--10 keV non-thermal luminosities obtained via the PL assumption (with $\Gamma=1.7$) for the underlying spectra, and the down-pointing triangles shows the 0.3--2.0 keV thermal luminosities obtained via BB assumption (with kT = 300 eV) (see discussion on \S\ref{ss:sa}). Spectrum-derived PL flux for J1831--0952 is shown with green triangle. $\dot{E}$ values are corrected for the Shklovskii effect where relevant \citep{Shklovskii1970}.}
         \label{fig:xrayeff1}
   \end{figure*}

Assuming temporarily that the X-ray fluxes of our candidates are solely attributed to the pulsars and that the distance estimates are close to reality, we show in Figure \ref{fig:xrayeff1} the $L_{X}$ vs $\dot{E}$ for the survey and archival sources for both thermal and non-thermal emission. 
Many sources in our sample cluster around an efficiency $\eta\sim 10^{-4}$. Only 3 pulsars from our list of archival X-ray observations have  parallactic distances (highlighted with yellow in Figure \ref{fig:xrayeff1}), i.e., reasonably firm flux estimates.  Except for J0711--6830, the two DM-distance estimates agree within a factor of 2 for the rest of the pulsars.

The solitary ms-pulsar J0711--6830  has an X-ray efficiency that is 1-1.5 orders of magnitude lower than the typical value. However, if the distance uncertainty is taken into account, this X-ray efficiency is very uncertain. The source does not have a parallax measurement and there is factor of 8 difference between the two DM-based distance ($(D_2/D_1)^2 = 61.1$).  If we assume the pulsar is located at 0.86 kpc based on the \emph{NE2001} model, we obtain a luminosity of $L_{0.3-2\ \mathrm{keV}} = 5.2 \times10^{29}$~\lumcgs for the BB and $L_{1-10\ \mathrm{keV}} = 2.5\times10^{30}$~\lumcgs for the PL model which places the source at the typical $10^{-4}$ range in the X-ray efficiency diagram.

Another outlier in Figure \ref{fig:xrayeff1} is the middle-aged pulsar J1926--1314 with an X-ray efficiency of $ \eta_{\rm X} \sim 10^{-1.5}$. 
We note that the X-ray efficiency only marginally changes if the \emph{NE2001}-based distance is used.
As mentioned in \S\ref{sec:hiddenp}, this source is thought to have a factor 10 stronger dipole magnetic field than a typical middle-aged pulsar. If the X-ray source is indeed the pulsar counterpart and the DM-distances are correct, this could indicate a case of magnetothermal heating (see \S\ref{ssJ1926}).

As it can be seen from Figure \ref{fig:xrayeff1}, our X-ray efficiency values do not vary much for different spectra and energy bands. One reason is that we converted a fixed number of counts in the same energy band to fluxes for both spectral models. Therefore, a large flux variation is not expected. More reliable X-ray efficiencies can be obtained in future works if pulsars have better distance estimates and their thermal vs non-thermal components are resolved with follow-up observations as illustrated by PSR\,J1831--0952.

\section{Summary and conclusion}

   \begin{enumerate}
   \setlength\itemsep{0.5em}
      \item We searched for X-ray counterparts for 14 nearby pulsars as part of our \emph{XMM-Newton} survey. We analyzed data in all EPIC cameras and obtained background-subtracted source counts and $3\sigma$ upper limits for non-detections.
      
      \item We detected two X-ray counterpart candidates at the pulsar positions with over $\sim 4.3 \sigma$ significance and one candidate with $\sim 3.5\sigma$ significance.
     
     \item We report on 8 X-ray counterpart candidates
     in the 4XMM-DR10 catalogue for which we could not find a respective note on the X-ray detection of the pulsar in the literature.

     \item We suspect that some of our initially chosen distance
     values may have been underestimated.
     For the optimized continuation of our survey, a more promising approach is to compare different distance models and choose only pulsars with good parallactic distances or small differences of DM-based distances.
     
     \item Assuming that the X-ray sources are indeed the pulsar counterparts, we calculated the X-ray fluxes, X-ray luminosities, and X-ray efficiencies of these pulsars with two stand-in model spectra, and provided  upper limits for our undetected pulsars.

      \item We assessed the possibility of an alternative MW counterpart of the reported  X-ray sources using their  $3\sigma$ position uncertainties in combination with the \emph{Gaia}-$\rm{EDR3}$, $\rm{2MASS}$, and $\rm{TESS}$ source catalogs.
      
  \item We argued the possibility of extended emission in PSR J1831--0952 by examining both \emph{XMM-Newton} and \emph{Chandra} data.
  
   \item We speculated that the high X-ray efficiency of PSR J1926--1314 could be attributed to the magneto-thermal heating if the distance is correct and the three stellar sources are excluded as counterparts of or contributors to the X-ray source in the pulsar vicinity.
   \end{enumerate}

\begin{acknowledgements}
      We thank the anonymous referee for constructive comments that helped to improve the manuscript. This work was supported by the Bundesministerium f{\"u}r Wirtschaft und Energie through Deutsches Zentrum f{\"u}r Luft-und Raumfahrt (DLR) under the grant number 50 OR 1917. GGP acknowledges support from the ACIS Instrument Team contract SV4-74018 issued by the \emph{Chandra} X-ray Observatory Center, which is operated by the Smithsonian Astrophysical Observatory for and on behalf of NASA under contract NAS8-03060. This research has made use of the VizieR catalogue access tool, CDS, Strasbourg, France (DOI : 10.26093/cds/vizier). The original description of the VizieR service was published in 2000, A\&AS 143, 23.
\end{acknowledgements}

\textit{\large{Software}}: \texttt{Astropy} \citep{astropy:2018}, \texttt{NumPy} \citep{harris2020array}, \texttt{Matplotlib} \citep{Hunter:2007}, \texttt{Seaborn} \citep{Waskom2021}

%
%
\bibliographystyle{aa}

\end{document}